\documentclass[runningheads]{llncs}
\usepackage[hidelinks]{hyperref}
\usepackage[T1]{fontenc}
\usepackage{booktabs}
\usepackage{caption}
\usepackage{multirow}
\usepackage{subcaption}
\usepackage{siunitx}
\usepackage{tabularx}
\usepackage{float}  
\usepackage{xcolor}
\usepackage{seqsplit}
\usepackage{booktabs}
\usepackage{orcidlink}
\usepackage[misc,geometry]{ifsym}
\usepackage{orcidlink}

\usepackage{graphicx}
\begin{document}
\title{Investigating Students’ Preferences for AI Roles in Mathematical Modelling: Evidence from a Randomized Controlled Trial}
%
%
\author{
Wangda Zhu\textsuperscript{1}\orcidlink{0000-0001-9611-4800} \and Guang CHEN\textsuperscript{2}\orcidlink{0009-0000-2738-4129} \and Yumeng Zhu\inst{3}\orcidlink{0000-0003-1632-9954} \and Lei Cai\inst{4}\orcidlink{0009-0007-1133-8639} \and Xiangen Hu\inst{5}\orcidlink{0000-0001-9045-4070}}
\authorrunning{Wangda et al.}
%
\institute{School of Design, The Hong Kong Polytechnic University, Hong Kong SAR.\and Department of Computing, The Hong Kong Polytechnic University, Hong Kong SAR. \and
College of Education, Zhejiang University, Hangzhou, China. \and College of Computer Science and Technology, Zhejiang University, Hangzhou, China. \and Department of Applied Social Sciences, The Hong Kong Polytechnic University, Hong Kong SAR.}
\maketitle              
\begin{abstract}
Mathematical modelling (MM) is a key competency for solving complex real-world problems, yet many students struggle with abstraction, representation, and iterative reasoning. Artificial intelligence (AI) has been proposed as a support for higher-order thinking, but its role in MM education is still underexplored. This study examines the relationships among students’ design thinking (DT), computational thinking (CT), and mathematical modelling self-efficacy (MMSE), and investigates their preferences for different AI roles during the modelling process. Using a randomized controlled trial, we identify significant connections among DT, CT, and MMSE, and reveal distinct patterns in students’ preferred AI roles, including AI as a tutor (providing explanations and feedback), AI as a tool (assisting with calculations and representations), AI as a collaborator (suggesting strategies and co-creating models), and AI as a peer (offering encouragement and fostering reflection). Differences across learner profiles highlight how students’ dispositions shape their expectations for AI. These findings advance understanding of AI-supported MM and provide design implications for adaptive, learner-centered systems.

\end{abstract}

\section{Introduction}
Mathematical modelling (MM) has become a cornerstone of contemporary mathematics education, reflecting its importance in preparing students to engage with complex real-world challenges such as climate change, pandemics, and technological innovation. International curricula and standards emphasize modelling as a key competency. For example, the Common Core State Standards for Mathematics (CCSSM) in the United States highlight modelling as one of the Standards for Mathematical Practice, requiring students to use mathematics as a tool for interpretation, representation, and decision-making. Similarly, the Programme for International Student Assessment (PISA) positions modelling as a central indicator of mathematical literacy, with Level 5 and 6 performers expected to model complex situations and reason flexibly across contexts. Yet PISA 2022 data show that only 27\% of Hong Kong’s 15-year-old students reached these top levels, suggesting that a majority continue to face challenges in modelling competence despite curricular emphasis. Since 2017, the Hong Kong Education Bureau has formally incorporated modelling tasks into the secondary curriculum, but many teachers have reported difficulties in both understanding and teaching MM effectively, limiting students’ opportunities to develop this essential skill.

Alongside these educational developments, artificial intelligence (AI) has emerged as a transformative force in learning environments. AI is increasingly used to provide adaptive feedback, scaffold cognitive processes, and create interactive environments that support higher-order thinking. In mathematics education, AI has been studied as a tutor, tool, or learning companion, offering ways to personalize instruction and expand students’ problem-solving repertoires. These developments raise a pressing question: what role should AI play in supporting mathematical modelling, which requires not only technical accuracy but also creative framing, abstraction, and iteration?

Despite its potential, research on the role of AI in MM education remains limited. Existing studies focus mainly on general mathematics problem solving or automated feedback systems, leaving a gap in understanding how AI can specifically support students during the iterative and complex process of modelling. Moreover, little is known about how learners with different profiles in design thinking (DT), computational thinking (CT), and mathematical modelling self-efficacy (MMSE) perceive AI’s role, or how these preferences should inform the design of adaptive systems. This is a critical oversight because MM requires both divergent and convergent thinking, and students approach modelling with diverse strengths and dispositions that AI could either support or hinder depending on its role.This study addresses these gaps through three research questions:
\begin{itemize}
    \item RQ1: What are the relationships among students’ design thinking (DT), computational thinking (CT), and mathematical modelling self-efficacy (MMSE)?
    \item RQ2: What are students’ preferences for AI roles during the mathematical modelling process?
    \item RQ3: How do groups of students with different levels of DT, CT, and MMSE vary in their preferences for AI roles?
\end{itemize}

By exploring these questions, this paper makes several contributions. First, it empirically establishes the connections among DT, CT, and MMSE, clarifying how these capacities interact in modelling contexts. Second, it provides novel insights into students’ preferences for AI roles, highlighting the types of support learners expect during MM processes. Third, it reveals how these preferences differ across learner profiles, offering practical guidance for designing AI systems that adapt to diverse student needs. Together, these contributions advance the theoretical understanding of MM education and inform the practical development of AI-supported learning environments.

\section{Literature Review}
\subsection{Mathematical modelling}

Mathematics is the science that emphasizes the formation of thinking abilities \cite{samo2017developing}. Math learning is important and fosters students’ high-order thinking skills, such as problem-solving skills, critical thinking, design thinking, and computational thinking skills \cite{gradini2025fostering,maslihah2020role}. However, pure mathematics is always abstract and distant from students’ real-life experiences. Mathematical modelling bridges this gap by connecting abstract mathematical concepts with students’ everyday contexts \cite{blomhoj2009different}. Modelling or model building means “\textit{the entire process leading from the original real problem situation to a mathematical model}”, which is “\textit{the most important part of the process of relating mathematics to the real world.}”\cite{blum1991applied}  Mathematical modelling can be described as an activity that involves transitioning iteratively between reality and mathematics. Teaching mathematical modelling is regarded as a form of reality-based teaching in mathematics education \cite{wess2021measuring}. This reality-based teaching has proven to be a valuable method \cite{maass2019role,wei2022can}, and has been identified as a powerful approach for developing mathematical understanding and higher-order thinking skills \cite{english2005mathematical,samo2017developing,wei2022can}. \\

The learning of mathematical modelling has become a key competency within school curricula and educational standards in many countries of the world \cite{arseven2015mathematical,cevikbas2022systematic}. For example, according to the Common Core State Standards for Mathematics (CCSSM), which describes varieties of expertise that mathematics educators at all levels should seek to develop, students should be able to model with mathematics and use appropriate tools strategically to solve problems \cite{ccsso2022common}. These mentioned competencies echo with the definition of high-order thinking skills, including computational thinking skills, design thinking. For example, establishing relationships and identifying patterns, and building a descriptive and representative model are core competencies mentioned in computational thinking skills \cite{barcelos2018mathematics}. Similarly, design thinking emphasizes iterative processes of empathizing, defining, ideating, prototyping, and testing, which parallel the stages of mathematical modeling where real-world problems are framed, translated into mathematical terms, and refined through evaluation \cite{brown2009change}. Thus, mathematical modeling can be understood as a central practice that integrates computational and design thinking, fostering epistemic fluency that enables students to move flexibly across different modes of reasoning and representation. \\

\subsection{Teaching and learning strategies for mathematical modelling }
Mathematical modelling tasks have inherent cognitive complexity. This means they demand certain competencies from students. Specifically, students need to be able to transfer knowledge from reality to mathematics \cite{blum2009mathematical,jablonski2024challenges}. Moving from informal reasoning to formal mathematics is both challenging for students to understand and critical for teaching \cite{pattison2016mathematics}. Students may have trouble figuring out what information to include, how to organize it, and how to turn qualitative relationships into quantitative ones \cite{blum2015quality}. These challenges necessitate an understanding of students’ modelling processes and the design of pedagogical strategies to support different types of learners. \\

To understand individual modelling processes, researchers have proposed various theoretical frameworks. In a typical mathematical modelling cycle, students interpret a real-world situation, abstract the essential elements, construct a \textit{situation model}, refine it into a \textit{real model}, and finally formalize it as a \textit{mathematical model} that can be analyzed and validated \cite{jablonski2024challenges,leiss2010role}. Specifically, constructing a real model involves simplifying and structuring a given real-world situation according to the problem solver's knowledge and interests. This process leads to problem formulation and the development of a real model of the situation. Kaiser and Sriraman (2006) identified all essential cognitive processes involved in solving modelling tasks, emphasizing the "situation model" and the "real model" as particularly important elements \cite{kaiser2006global}. Building on Blum and Leiss's (2002) modelling cycle, Leiss et al. (2010) further validated the importance of the situation model \cite{leiss2010role}, which is a personal cognitive structure that reflects one's understanding of a situation, serving as the mental counterpart to the situation structure \cite{blum2002icmi}. \\ 

Learners approach mathematical modelling with diverse profiles, which requires differentiated instructional strategies. Students vary in their cognitive strengths. Some excel in logical-symbolic reasoning, while others are more oriented toward spatial or interpersonal skills. This suggests that mathematics instruction should leverage these differences through visual representations, hands-on activities, or collaborative work. An approach that  aligns with Gardner’s theory of multiple intelligences \cite{gouws2011teaching}. Furthermore, novices tend to be passive, giving up or waiting for help when stuck, whereas expert modellers see modelling as a collaborative, sense-making process \cite{drakesNoDatePeter}.\\ 

Pedagogically, teachers are encouraged to provide students with more competency - oriented activities, foster connections across concepts, prompt regular reflection, and maintain a balanced interplay between independent and teacher-directed learning phases \cite{blum2007investigating}. A growing body of research shows that teachers’ guidance and social interaction can improve students’ learning outcomes in mathematical modelling \cite{arseven2015mathematical,ferri2018learning}. For example, Gürel (2023) found that the teacher who acted as a monitor provided higher support levels compared to the teacher who acted as a partner\cite{gurel2023teaching}. However, most teachers lack extensive experience in mathematical modelling themselves. Moreover, classroom constraints limit the extent to which each student can receive personalized, real-time support across the full modelling cycle. These limitations highlight the need for complementary instructional approaches.\\

\subsection{AI in mathematical learning}

Large language models (LLMs) create new opportunities for tailored, immediate feedback and genuinely personalized learning. They can rapidly process complex problems, including mathematical reasoning and the production of tightly structured argumentation, capabilities that align well with the demands of mathematics education and have been validated across multiple studies \cite{imani2023mathprompter,blair2023can1}. For example, Luong et al. leveraged AI’s capacity to personalize instructional content to strengthen students’ mathematical skills \cite{luong2025personalized}, and Noviyana et al. reported that creatively designed, AI-supported pedagogical patterns improved mathematics learning outcomes \cite{noviyana2025enhancing}.\\

Beyond establishing effectiveness, it is also critical to understand how AI shapes the learning process. Jin et al. found that AI involvement can reduce learners’ cognitive demand, thereby improving study efficiency \cite{jin2025exploring}. In addition to these cognitive effects, Van et al. showed that AI can spark creativity and curiosity, motivating students to engage in self-directed exploration of mathematical problems \cite{van2023teaching}. 
This kind of generative, student-led pedagogy is highly effective in primary education \cite{zhu2025bridging}.
Extending this evidence base, Deng et al. (2025) conducted a meta-analysis of 62 studies on ChatGPT in education and concluded that it can meaningfully enhance learning performance while improving affective–motivational facets of learning \cite{deng2025does}. Taken together, these findings indicate that AI participation can simultaneously ease mental workload and stimulate interest and creative thinking, supporting deeper inquiry~\cite{zhu2024enhancing}.\\

However, the manner in which AI is used matters: unreflective adoption may compromise the learning process \cite{carvalho2022can}. At the cognitive level, inappropriate reliance can lead learners to bypass genuine cognitive engagement \cite{nathanpotency}. A large-scale field study further shows that when students use GPT-4 without constraints during practice, short-term gains may mask longer-term costs: once assistance is removed, their independent problem-solving performance can fall below that of peers who never used AI \cite{bastani2025generative}. On the affective side, poorly designed interactions with LLMs may have negative emotional consequences; when the system lacks sufficient emotional intelligence, students often fail to experience resonance or encouragement due to the absence of humanized interaction \cite{birenbaum2023chatbots,park2024promise1}.\\

These cognitive and affective considerations suggest that generic, general LLM interaction methods should not be transplanted directly into educational contexts, especially in mathematics modelling, which places high demands on both cognition and emotion. To address this challenge, the present work adopts the “AI role” approach recommended by Wang et al. \cite{wang2025effect}. This strategy not only affords interactions that feel more human-like and affectively supportive, but also enables regulation of learners’ cognitive load through deliberate variation of AI roles.\\

\subsection{ AI roles in mathematical learning}
An AI role, or AI character, refers to the functional persona that an LLM adopts during interaction, which shapes what kind of help is offered, when it is provided, and in what form \cite{pataranutaporn2021ai}. In traditional classrooms, educators frequently shift between roles (instructor, facilitator, observer), and even encourage students to assume teaching roles, as such variation can deepen engagement and understanding. For instance, having students explain concepts to others is known to foster active cognitive processing and motivation, a method often referred to as Learning by Teaching (LBT) \cite{frager1970learning}. This principle extends naturally to AI-assisted learning: a large language model can likewise adopt different roles to influence the learning process \cite{lin2025enhancing,song2024students}. Jin et al. employed an LBT-inspired approach by alternating the AI between student and questioner roles, and found that this mode shifting stimulated more knowledge-dense dialogue, illustrating the feasibility of transferring role-shifting mechanisms into AI-mediated learning \cite{jin2024teach}.\\

Different roles may exert distinct influences on learners. Bailing et al. investigated five AI agents with different personalities and examined their effects on learning outcomes, as well as cognitive and affective factors. They found that personality differences indeed produced divergent impacts; for instance, agents emphasizing openness and conscientiousness facilitated deeper cognitive processing such as explanation \cite{lyu2025role}. Their follow-up work further showed that personality style shaped learner behavior, with some personalities eliciting more cognitive support while others triggered more non-responsive behaviors \cite{lyu2025exploring}. Beyond personality, the very nature of the role assumed by the AI can also affect learning. Learners may project AI-mediated interactions onto traditional classroom contexts \cite{baylor2005simulating}, perceiving greater cognitive pressure when the AI plays a teacher role, while experiencing lighter and more enjoyable interactions when the AI adopts a student role \cite{li2025facts}.\\

In mathematical modelling, learners must engage in substantial recognitive elaboration, a requirement that may favor AI in the teacher role. Kestin et al., through a randomized controlled trial, compared a customized GPT-4 tutor with traditional classroom instruction, and found that students supported by the AI tutor achieved significantly higher scores in less time, while also reporting greater engagement and motivation \cite{kestin2025ai}. Similarly, Kumar et al. demonstrated that AI dialogues framed explicitly in a teacher role produced stronger learning outcomes than role-neutral interactions \cite{kumarmath}. Nonetheless, an overly strict teacher identity can sometimes impose excessive pressure, potentially evoking negative emotional experiences that hinder learning.\\

An alternative approach is to position AI as a learner or peer. This configuration often creates a relaxed, low-pressure environment, in which students feel more comfortable “teaching” the AI. In turn, this encourages explanation and reflection, thereby promoting active learning. Xing et al. designed the generative AI–powered teachable agent “ALTER-Math,” enabling middle school students to teach the AI as part of their mathematics learning. Across three stages of study, they demonstrated significant cognitive gains for students using the AI learning companion \cite{xing2025development1}. Likewise, Chen et al. explored the use of ChatGPT as a teachable agent in programming education, finding that dialogic teaching with the AI promoted knowledge internalization. Learners also reported stronger self-efficacy and greater use of self-regulated learning strategies, suggesting that teaching-oriented AI companions can enhance both motivation and learning regulation \cite{chen2024learning1}. In general, the student role offers distinct advantages in fostering emotional engagement and supportive learning environments.\\

Current research has provided substantial evidence of the effects of AI in either a teacher or a student role on learners. However, direct comparisons between different roles remain limited. Therefore, the central question of this study is whether distinct AI roles lead to measurable differences in students’ learning outcomes and their learning experience, particularly in terms of creativity self. By examining AI as a tutor versus AI as three different types of students, we aim to determine how the role assumed by an AI partner can simultaneously support deeper learning and nurture positive emotional experiences in mathematical problem-solving contexts.
\section{Methods}
We employed a within-subjects, counterbalanced experimental design to develop and evaluate appropriate AI roles for students in mathematics modelling learning. Five traditional in-class roles (Tutor, Teaching Assistant, Peer Student, Struggling Student, and Excellent Student) served as the primary focus. In addition, a baseline condition without any assistance was included for comparison prior to introducing the five roles.
In each condition, the AI took on one role and guided students through the same difficult-level mathematical modelling problem. This ensured that differences in outcomes could be attributed to the role rather than task variation. The design also allowed us systematically examine how each AI role influenced students’ performance and perceptions under comparable conditions.

\begin{table}[t]
\centering
\caption{Experimental Design: Group Assignment and AI Role Rotation Across Five Rounds}
\label{tab:experimental_design}
\begin{tabular}{lccccccc}
\toprule
\textbf{Group} & \textbf{N} & \textbf{Baseline} & \textbf{Round 1} & \textbf{Round 2} & \textbf{Round 3} & \textbf{Round 4} & \textbf{Round 5} \\
 & & & (Task A) & (Task B) & (Task C) & (Task D) & (Task E) \\
\midrule
G1 & 6 & No assistance & Tutor & TA & Peer & Struggling & Excellent \\
G2 & 6 & No assistance & Excellent & Tutor & TA & Peer & Struggling \\
G3 & 6 & No assistance & Struggling & Excellent & Tutor & TA & Peer \\
G4 & 6 & No assistance & Peer & Struggling & Excellent & Tutor & TA \\
G5 & 6 & No assistance & TA & Peer & Struggling & Excellent & Tutor \\
\bottomrule
\end{tabular}
\end{table}

\subsection{Study Design}
\subsubsection{Participants}
A total of 65 individuals were recruited through social media. Of these, 32 did not participate in the experiment, 3 withdrew during the experimental session, and 4 were excluded due to anomalous or invalid data in the online task. The final analytical sample consisted of 26 participants. All participants were university students aged 18 or older, enrolled in non-STEM majors, and had not taken any university-level mathematics courses. Their highest level of mathematics education was at the secondary school level, ensuring their mathematical proficiency was comparable to that of typical high school graduates.

The study employed a within-subjects, counterbalanced experimental design: each participant completed a baseline task (no assistance) followed by five tasks with an AI agent assuming a different role in each round. Participants were randomly assigned to one of five counterbalanced groups (N = 6 per group), with AI role sequences rotated across groups as shown in Table~\ref{tab:experimental_design}. Due to the unpredictability of online recruitment, the final sample size and composition differed slightly from the initial plan. Descriptive statistics for the recruited participants, including gender distribution, experimental group allocation, and age, are summarized in Table \ref{tab:combined_table}.

\begin{table}[t]
\centering
\caption{Demographic Characteristics and Age Statistics of the Final Sample ($N = 26$)}
\label{tab:combined_table}
\resizebox{\textwidth}{!}{%
\begin{tabular}{llcccc}
\toprule
\textbf{Variable} & \textbf{Category / Statistic} & \textbf{Female} & \textbf{Male} & \textbf{Non-binary} & \textbf{Total} \\
\midrule
\multirow{3}{*}{Gender (Frequency \%)} 
    & Female & 19 (73.1\%) &  &  & 19 (73.1\%) \\
    & Male &  & 6 (23.1\%) &  & 6 (23.1\%) \\
    & Non-binary / Other &  &  & 1 (3.8\%) & 1 (3.8\%) \\
\hline
\multirow{5}{*}{Experimental Group (Frequency \%)} 
    & G1 & \multicolumn{3}{c}{} & 4 (15.4\%) \\
    & G2 & \multicolumn{3}{c}{} & 6 (23.1\%) \\
    & G3 & \multicolumn{3}{c}{} & 5 (19.2\%) \\
    & G4 & \multicolumn{3}{c}{} & 5 (19.2\%) \\
    & G5 & \multicolumn{3}{c}{} & 6 (23.1\%) \\
\hline
\multirow{1}{*}{Age Statistics} 
    & Mean age $\pm$ SD & 22.3 $\pm$ 2.0 & 20.8 $\pm$ 1.8 & 25.0 & 22.0 $\pm$ 2.0 \\
\bottomrule
\end{tabular}
}
\end{table}

\subsubsection{Materials}
To obtain both process and perception data, several types of materials were used throughout the experiment. These included (a) complete chat logs from each task to document participants’ interactions with the AI agent, (b) a pre-task questionnaire to assess baseline competencies, (c) short post-task questionnaires to capture immediate perceptions of each AI role, and (d) a final questionnaire to collect retrospective evaluations. Each type of material is described below.

\paragraph{Chat Logs}
All written solutions from the baseline task (completed without assistance) and the full chat logs of all interactions with the AI agent across the five experimental rounds were collected. These logs include every participant input, AI response, and dialogue turn, providing a detailed record of each problem-solving process for later analysis.

\paragraph{Pre-task Questionnaire}
The pre-task questionnaire measured participants’ baseline levels of mathematical modelling self-efficacy, computational thinking, and design thinking. All items were administered in English, with a Chinese translation shown in smaller font to aid comprehension without altering the original wording.

\begin{itemize}
    \item Mathematical modelling self-efficacy(MMSE) was assessed using the 17-item Mathematical Modelling Self-Efficacy Scale~\cite{koyuncu2016development}. Participants rated their agreement with statements on a 5-point Likert scale (1 = strongly disagree, 5 = strongly agree). A higher score reflects stronger belief in one’s ability to engage in mathematical modelling.
    \item Computational thinking(CT) was assessed using the 29-item Computational Thinking Scale~\cite{korkmaz2017validity}. The scale measures five dimensions: Creativity, Algorithmic Thinking, Cooperativity, Critical Thinking, and Problem Solving. Responses are rated on a 5-point Likert scale (1 = never, 5 = always).
    \item Design thinking (DT) traits were assessed using a 9-item scale~\cite{blizzard2015using}. The scale measures five dimensions: feedback seeking, integrative thinking, optimism, experimentalism, and collaboration. Responses use a 0–4 format (0 = strongly disagree, 4 = strongly agree) as in the original instrument; all scores were retained in their original metric for analysis.
\end{itemize}

\paragraph{Post-task Questionnaire}\label{session:Post-task Questionnaire}
After each experimental round, participants completed a short questionnaire about their experience with the AI role. It included four closed-ended items rated on Likert scales and one open-ended question inviting comments. The complete set of items is shown in Appendix~\ref{app:questionnaires}.

\paragraph{Overall Questionnaire}\label{session:Overall Questionnaire}
After completing all five tasks, participants completed a final retrospective questionnaire evaluating their overall experience with the AI roles. It included (1) a ranking of the five AI roles from most to least preferred and (2) an open-ended question about perceived differences. The wording of these items is provided in Appendix~\ref{app:questionnaires}.

\subsubsection{Experimental Procedure}
The full study was conducted entirely online and consisted of three phases: a pre-session, a task session, and a post-session.

\paragraph{Pre-Session (approximately 5 minutes)}
Participants contacted the assigned helper through the provided communication channel prior to the experiment. The helper confirmed participants’ identity and relevant information. The helper then provided the necessary experimental materials and guided participants through signing the informed consent form. After signing the consent, the helper gave participants a weblink to access the pre-task questionnaires. Participants completed three digital surveys assessing baseline competencies: the Mathematical Modelling Self-Efficacy Scale, the Computational Thinking Scale, and the Design Thinking Scale.

\paragraph{Task Session (approximately 30 minutes)}
After completing the three pre-task questionnaires, the web interface automatically redirected participants to the main task session. First, the webpage provided an explanation of the key concepts in mathematical modelling: Situation Model, Real Model, and Mathematical Model. An example problem was also presented, demonstrating how to solve a task following these three stages.

Participants then completed six short mathematical modelling tasks on a laptop via the experiment website. Each task lasted four minutes. The first task was completed without AI assistance (baseline), allowing participants to independently solve a mathematical modelling problem and gain a clear understanding of the main task for the experiment. This was followed by five tasks in which an AI agent assumed a distinct role for each round. After each of these five tasks, participants completed the post-task questionnaire described in Section \ref{session:Post-task Questionnaire}. Upon completing all six tasks, participants filled out the overall questionnaire described in Section \ref{session:Overall Questionnaire}. The task order and AI role assignment followed the counterbalanced design described in Table \ref{tab:experimental_design}.

\paragraph{Post-Session (approximately 5 minutes)}
After completing all tasks and the final questionnaire, participants contacted the assigned helper via the provided communication channel. The helper verified that the session was complete and then issued a compensation of approximately 7 USD in the form of an electronic supermarket voucher.

\subsubsection{Data Analysis}
Data analysis was conducted on a Windows 10 system using Python 3.12.7. Both quantitative and qualitative data were analyzed. Quantitative analyses included questionnaire responses, while qualitative analyses focused on participants’ open-ended responses regarding their perceptions of the AI roles.

\paragraph{Data Preparation}
All participants who completed the entire experiment were retained in the dataset. No cases were excluded or removed. We considered that inattentive or inconsistent responses among participants who finished the full procedure reflect real classroom conditions, where some students may not fully engage with modelling tasks. For this reason, these data were included in the analyses.

In addition, our analysis focused on participants’ rankings of their preferences for the five AI roles. Because the overall rating of all AI roles (see Section \ref{session:Overall Questionnaire}) was collected only after all tasks had been completed, some participants might have forgotten which AI role corresponded to each task. Therefore, the comparative responses recorded immediately after each task (see Section \ref{session:Post-task Questionnaire}) were used as a reference to verify and adjust the final rankings. This procedure ensured that the rankings accurately reflected participants’ true preferences for each AI role.

\paragraph{Quantitative Analysis}
For the three pre-task questionnaires, Pearson correlation coefficients were calculated to examine the relationships among participants’ baseline competencies. In addition, a partial least squares structural equation modelling (PLS-SEM) analysis was conducted using SmartPLS v.4.1.1.5 to investigate how computational thinking and design thinking traits influenced participants’ mathematical modelling self-efficacy. Subscale scores from the computational thinking and design thinking questionnaires were used as measurement indicators for these latent constructs. The model was then refined iteratively following SmartPLS guidelines (e.g., examining indicator loadings and modification indices) to achieve satisfactory reliability and validity.

Questionnaire data and task interaction data were analyzed using non-parametric methods due to the within-subjects design and the ordinal nature of some measures. Specifically, the Friedman test was employed to examine differences across the five AI role conditions. When the Friedman test indicated a significant effect, post-hoc pairwise comparisons were conducted using the Wilcoxon signed-rank test. The baseline (no assistance) condition was excluded from these comparisons.

To compare how participants with different ability levels evaluated the five AI roles, the sample was split into two groups (upper 50\% and lower 50\%) based on their total scores in mathematical modelling self-efficacy, design thinking, or computational thinking. When ties occurred at the cutoff point, participants were ordered by their assigned identification number before grouping. This procedure ensured an exact 50/50 split between the high- and low-score groups and made the grouping fully reproducible. Subsequent nonparametric Mann–Whitney U tests were then conducted to test for significant differences between these groups in their evaluations of the five AI roles.

\paragraph{Qualitative Analysis}
To clarify the perceived strengths and weaknesses of each AI role, we conducted a qualitative analysis of participants’ open-ended responses collected in the post-task questionnaire (see Section \ref{session:Post-task Questionnaire}). First, all responses were read thoroughly to gain an overall understanding of the content. Preliminary themes were then identified in an open and exploratory manner. These themes were subsequently reviewed, compared, and combined where appropriate. Through this iterative process, clear and distinct categories were developed to represent recurring ideas and patterns in participants’ comments.

\subsection{Design and Development}
This study was designed to examine how different AI roles influence participants’ engagement and performance in mathematical modelling tasks. The design followed two main principles. First, it aimed to reflect real classroom conditions by using authentic problem-solving tasks and allowing normal differences in engagement. Second, it systematically varied the type of AI assistance so that each participant experienced every role, which allowed direct comparisons across roles while reducing individual differences.

\subsubsection{System Architecture}
The system uses a three-tier architecture with a presentation layer, application layer, and data layer. The presentation layer offered a browser-based interface. Through this interface, participants completed tasks and interacted with AI roles.
The application layer managed three key functions: task delivery, role-based AI response generation, and real-time chat interactions.
The data layer handled persistent storage of all participant responses and interaction logs. It also used reliability mechanisms to ensure data integrity.

A detailed technical description of the system implementation, including software stack, database structure, and communication protocols, is provided in Appendix~\ref{app:SystemArchitecture}.

\subsubsection{Content Design}
To help participants understand the experimental tasks, we included an additional task without AI assistance and a worked example. These were generated by GPT-5 based on textbook content.
The core of the experiment consisted of five primary mathematical modelling tasks, simplified to a high school level to ensure they could be completed within four minutes. 
To closely mirror the mathematical modelling learning process of high school students, we recruited university students who had not taken formal college-level mathematics courses. We selected this group because their mathematical proficiency remains comparable to that of high school students.

\subsubsection{AI Role Design}
To investigate optimal AI scaffolding in mathematical modelling education, we designed five distinct AI roles representing different support levels and pedagogical approaches:
\begin{itemize}
    \item Tutor: Offers expert-level guidance with systematic explanations and step-by-step solutions. This role actively teaches core concepts and guides participants through the three modelling stages.
    \item Teaching Assistant (TA): Offers medium to high support using inquiry-based scaffolding. The TA never gives direct answers. Instead, they help students discover answers through strategic questions and step-by-step hints.
    \item Peer Student: Acts as a collaborative learning partner with the same ability. It exchanges ideas, may make mistakes, and takes part in collaborative correction. This helps create an equal learning atmosphere.
    \item Struggling Student: Provides low or misleading support. This role shows confusion about key concepts, frequent errors, and uncertainty. It acts as a contrast to help examine the effect of low-quality peer modelling.
    \item Excellent Student: Provides correct answers quickly but with little explanation. Their responses are confident and concise, and they highlight the difference between answer-focused support and process-focused support.
\end{itemize}
All responses were generated with GPT-based models and role-specific system prompts. They were constrained to 100-150 characters to ensure ecological validity.
\section{Results}
\subsection{Mathematical Modelling Self-efficacy, Computational Thinking and Design Thinking}
\begin{figure}[t]
    \centering
    \includegraphics[width=\textwidth]{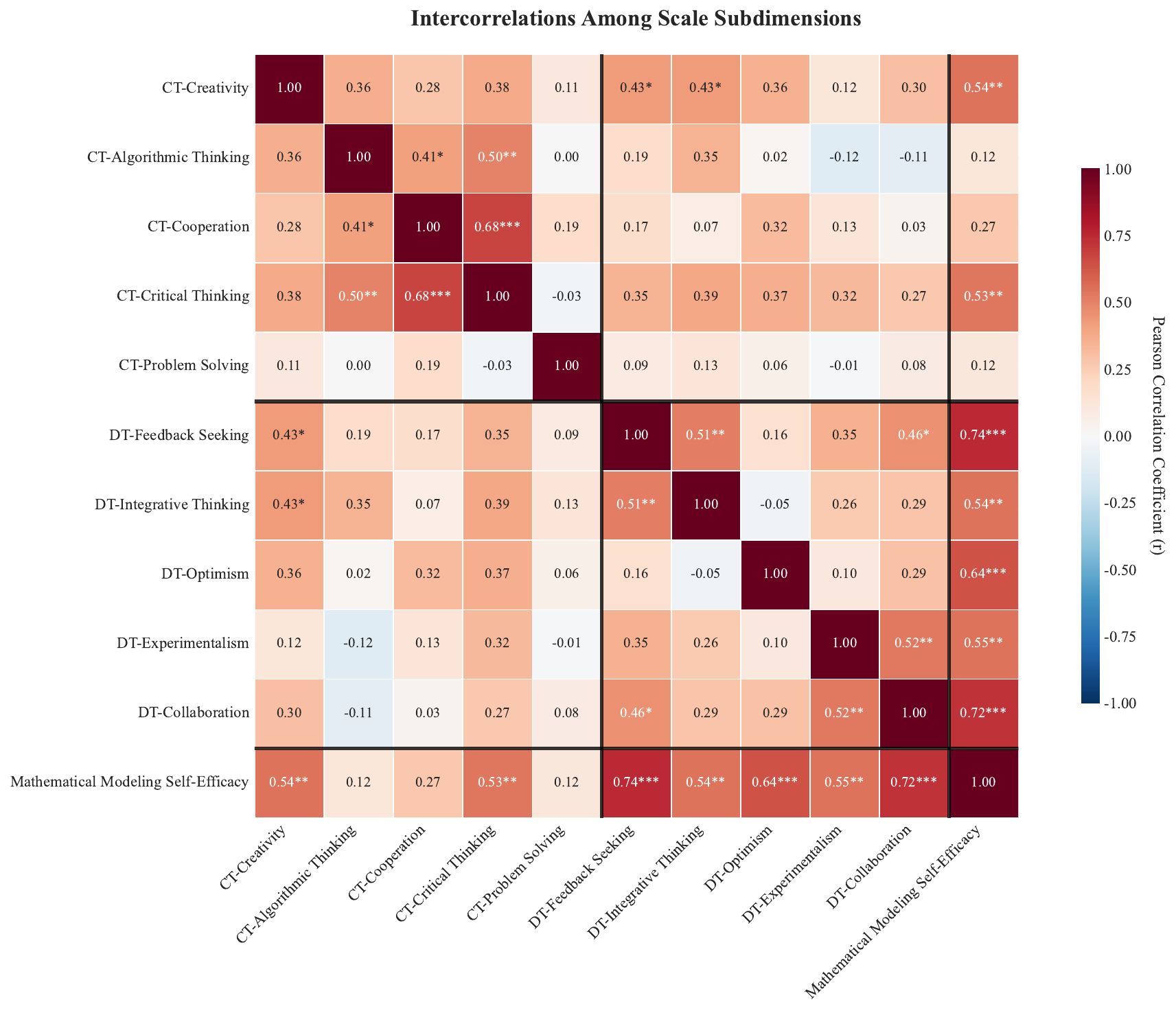}
    \caption{Intercorrelations among scale subdimensions. Pearson correlation coefficients (r) are presented, with asterisks indicating statistical significance: *$p < .05$, **$p < .01$, ***$p < .001$. The color gradient reflects the magnitude and direction of correlations, ranging from strong negative (blue) to strong positive (red).}
    \label{fig:correlation_matrix}
\end{figure}

This section examines the interrelationships among participants' mathematical modelling self-efficacy (MMSE), computational thinking (CT), and design thinking (DT). Figure~\ref{fig:correlation_matrix} presents the correlations among subdimensions of these constructs. Most correlations were positive, with several reaching statistical significance. This pattern indicates meaningful connections among the three competencies.

\begin{figure}[t]
    \centering
    \includegraphics[width=\textwidth]{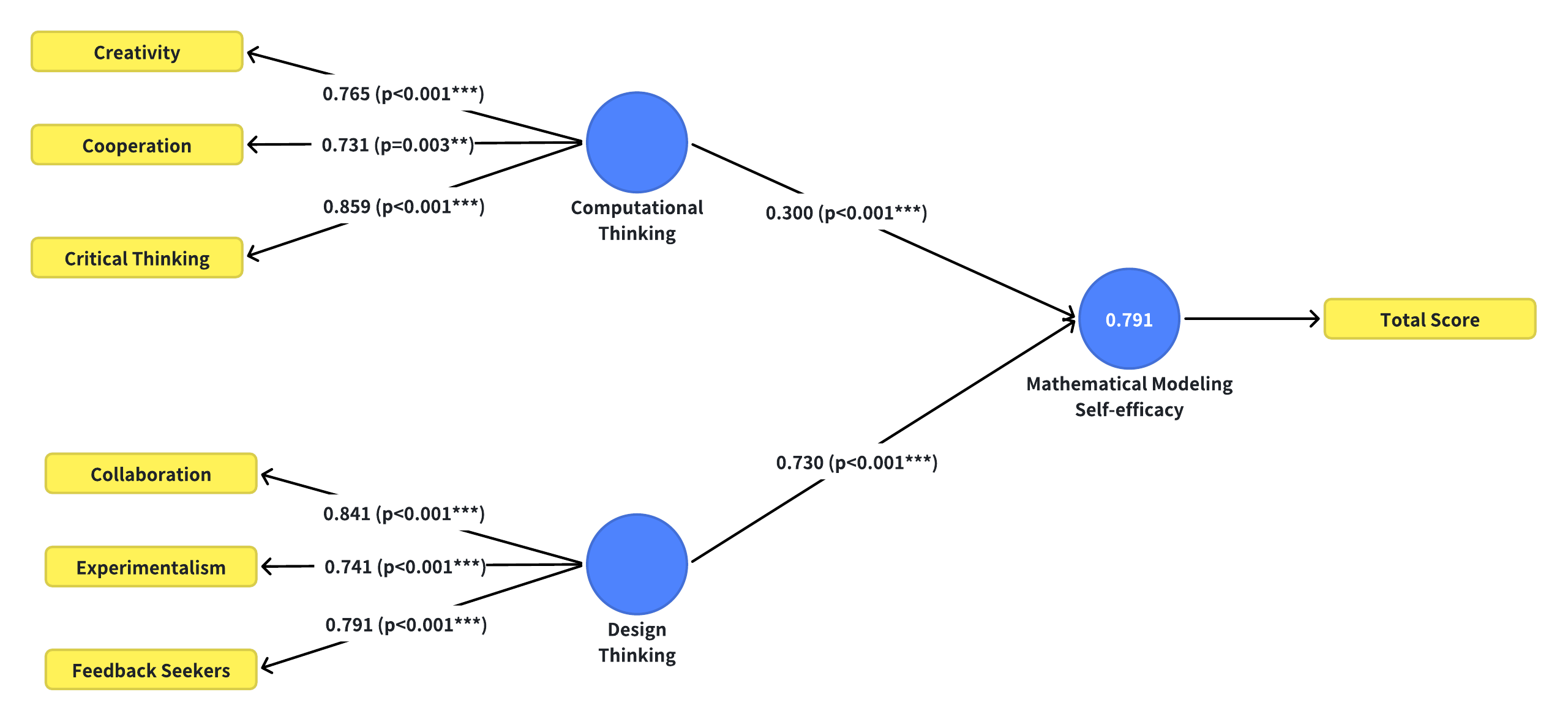}
    \caption{Structural equation model illustrating the relationships between Computational Thinking (CT), Design Thinking (DT), and Mathematical Modelling Self-efficacy (MM). The numbers represent standardized path coefficients, with asterisks indicating statistical significance (*$p < .05$, **$p < .01$, ***$p < .001$).}
    \label{fig:sem_model}
\end{figure}

To further investigate these relationships, we conducted partial least squares structural equation modelling (PLS-SEM). The resulting structural model is illustrated in Figure~\ref{fig:sem_model}. The measurement model demonstrated satisfactory reliability and validity (see Table~\ref{tab:measurement_model}). All outer loadings exceeded the recommended threshold of 0.70 \cite{hulland1999use}. Construct-level metrics also met established criteria: composite reliability (CR) values exceeded 0.60 \cite{fornell1981evaluating}, average variance extracted (AVE) values were above 0.50 \cite{fornell1981evaluating}, and variance inflation factor (VIF) values remained below 5.00 \cite{hair2011pls}. These results support the internal consistency, convergent validity, and absence of multicollinearity in the measurement model for both CT and DT constructs.

The structural model (Figure~\ref{fig:sem_model}) revealed that both DT ($\beta = 0.730$, $p < .001$) and CT ($\beta = 0.300$, $p < .001$) were significant positive predictors of MMSE. The model explained a substantial proportion of variance in MMSE ($R^2 = 0.808$, adjusted $R^2 = 0.791$). Effect sizes ($f^2$) were calculated to evaluate the magnitude of each predictor’s contribution: DT showed a very large effect ($f^2 = 2.288$), while CT showed a medium to large effect ($f^2 = 0.387$) on MMSE. 
Collectively, these results establish that both design thinking and computational thinking are significant contributors to mathematical modelling self-efficacy, with design thinking demonstrating a particularly strong predictive power.

\begin{table}[t]
\centering
\caption{Measurement Model Results: Reliability and Validity}
\label{tab:measurement_model}
\begin{tabular}{l c c c c c}
\toprule
\textbf{Construct / Indicator} & \textbf{Outer} & \textbf{Cronbach's} & \textbf{Composite} & \textbf{AVE} & \textbf{VIF} \\
& \textbf{Loading} & \textbf{Alpha} & \textbf{Reliability} & & \\
\midrule
\textbf{Computational Thinking (CT)} & & \textbf{0.705} & \textbf{0.829} & \textbf{0.619} & \\
\quad Creativity & 0.765 & & & & 1.167 \\
\quad Cooperation & 0.731 & & & & 1.843 \\
\quad Critical Thinking & 0.859 & & & & 1.981 \\
\addlinespace
\textbf{Design Thinking (DT)} & & \textbf{0.706} & \textbf{0.835} & \textbf{0.628} & \\
\quad Collaboration & 0.841 & & & & 1.560 \\
\quad Experimentalism & 0.741 & & & & 1.411 \\
\quad Feedback Seekers & 0.791 & & & & 1.291 \\
\bottomrule
\end{tabular}
\end{table}



\subsection{Ranking of AI Roles Preference Levels}
\subsubsection{Overall Ranking of AI Roles Preference}
A Friedman test was conducted to examine whether students showed differential preferences among five AI role personas (Tutor, TA, Peer, Struggling Student, and Excellent Student). The test revealed a significant effect of AI role on student rankings, $\chi^2(4, N = 26) = 28.83$, $p < .001$, indicating that students demonstrated distinct preferences for different AI pedagogical approaches.

\begin{figure}[t]
    \centering
    \includegraphics[width=\textwidth]{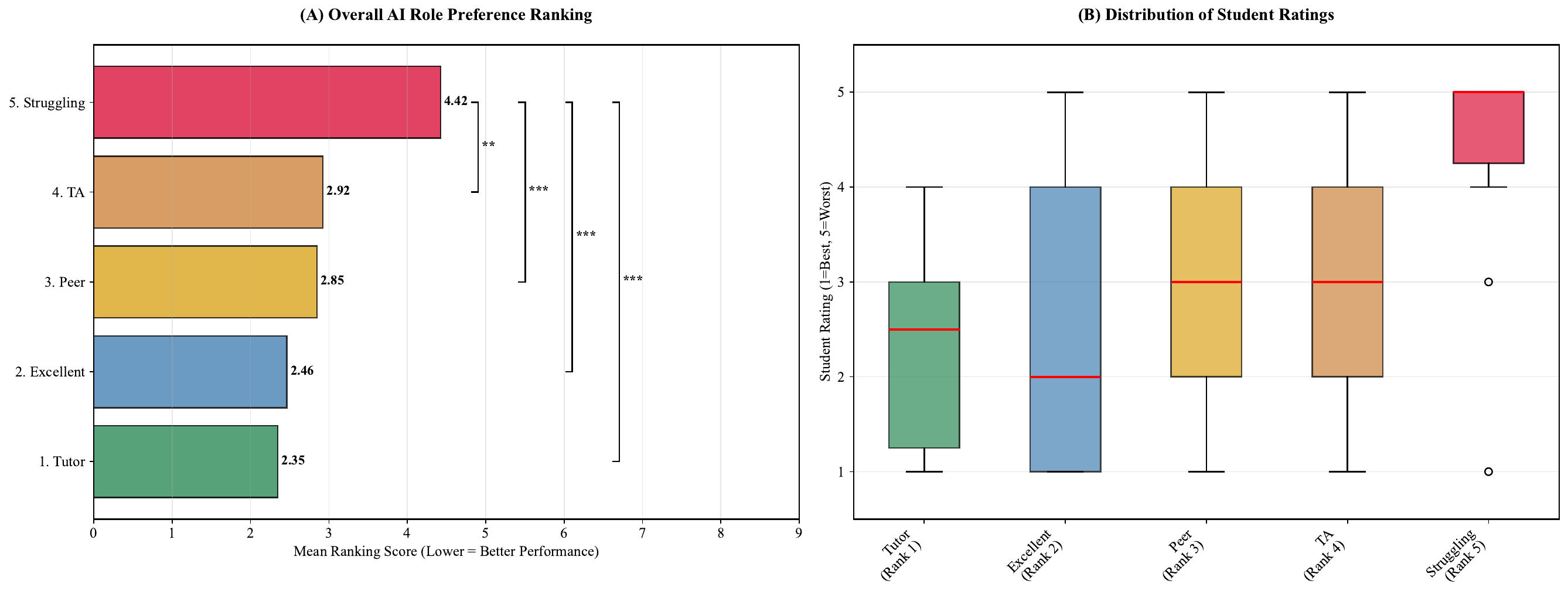} 
    \caption{A Participant preferences for AI role personas. Panel A displays mean ranking scores with standard errors for each AI role, where lower scores indicate greater preference. Brackets with asterisks denote statistically significant pairwise differences (** p < .01, *** p < .001, Bonferroni corrected). Panel B shows the distribution of individual student ratings (1 = best, 5 = worst) through box plots, with boxes representing interquartile ranges, red horizontal lines indicating medians, whiskers extending to minimum and maximum values, and circles representing outliers.}
    \label{fig:ai_role_rankings}
\end{figure}

Descriptive statistics for each AI role are presented in Table~\ref{tab:ai_rankings}. Rankings were made on a scale from 1 (most preferred) to 5 (least preferred). The Tutor role received the most favorable rankings ($M = 2.35$, $SD = 1.02$), followed by the Excellent Student role ($M = 2.46$, $SD = 1.30$), the Peer role ($M = 2.85$, $SD = 1.19$), and the TA role ($M = 2.92$, $SD = 1.41$). In stark contrast, the Struggling Student role was ranked least favorably ($M = 4.42$, $SD = 1.17$). The distribution of student ratings is illustrated in Figure~\ref{fig:ai_role_rankings}.

To identify specific differences among AI roles, post hoc pairwise comparisons were conducted using Wilcoxon signed-rank tests with Bonferroni correction ($\alpha = .005$). Results are summarized in Table~\ref{tab:pairwise_comparisons}. Four significant differences emerged, all involving the Struggling Student role. Specifically, the Struggling Student role was rated significantly less favorably than the Tutor role ($Z = 21.00$, $p < .001$, $r = 0.85$), the Excellent Student role ($Z = 28.00$, $p < .001$, $r = 0.80$), the Peer role ($Z = 47.50$, $p < .001$, $r = 0.67$), and the TA role ($Z = 64.00$, $p = .004$, $r = 0.57$). All effect sizes were large ($r > .50$), suggesting substantial preference differences. Importantly, no significant differences were found among the four constructive AI roles (Tutor, TA, Peer, and Excellent Student; all $ps > .14$), indicating that students valued competent and supportive AI personas similarly, regardless of the specific pedagogical approach embodied.

\subsubsection{Effects of AI Role on User Experience}

\begin{figure}[t]
\centering
\includegraphics[width=0.95\linewidth]{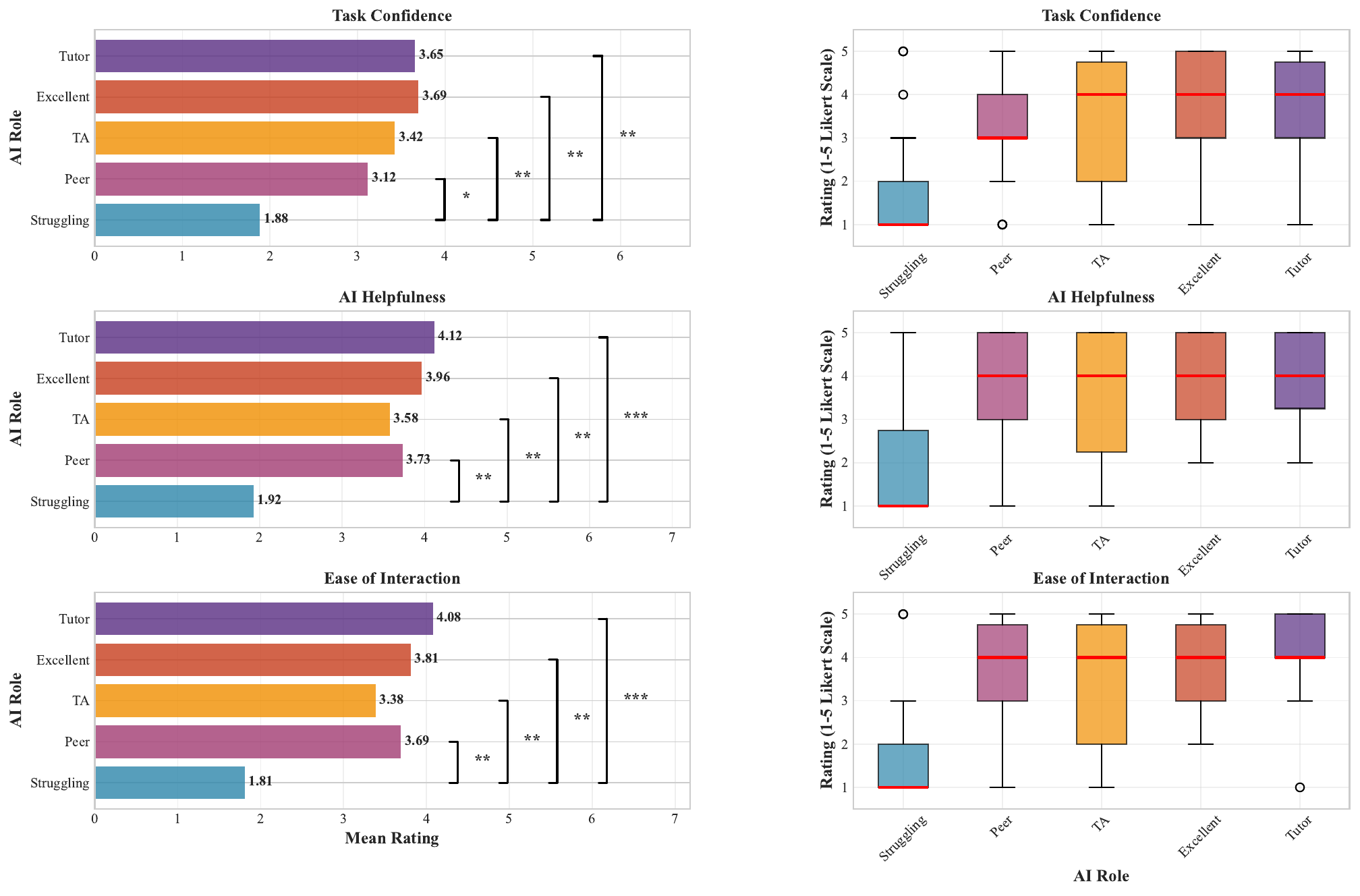}
\caption{
Perceived effectiveness of AI roles across three key dimensions (*$p < .05$, **$p < .01$, ***$p < .001$).
}
\label{fig:ai_role_significance}
\end{figure}

Three separate Friedman tests were conducted to examine whether AI role personas (Excellent Student, Peer, Struggling Student, TA, and Tutor) differentially affected students' task confidence, perceived AI helpfulness, and ease of interaction (see Figure~\ref{fig:ai_role_significance}). All 26 participants completed tasks with all five AI roles in a within-subjects design.

\paragraph{Task Confidence}

The Friedman test revealed a significant effect of AI role on task confidence, $\chi^{2}(4, N = 26) = 24.85$, $p < .001$. Descriptive statistics are presented in Table~\ref{tab:descriptive}. Students reported highest task confidence when working with the Excellent Student role ($M = 3.69$, $SD = 1.19$), followed by the Tutor role ($M = 3.65$, $SD = 1.16$), TA ($M = 3.42$, $SD = 1.39$), and Peer ($M = 3.12$, $SD = 1.14$). In contrast, the Struggling Student role elicited significantly lower task confidence ($M = 1.88$, $SD = 1.24$).

Post hoc pairwise comparisons using Wilcoxon signed-rank tests with Bonferroni correction ($\alpha = .005$) identified specific differences (see Table~\ref{tab:pairwise}). The Struggling Student role was associated with significantly lower task confidence compared to all four other roles: Excellent ($W = 28.5$, $p = .004$), Peer ($W = 31.0$, $p = .030$), TA ($W = 12.0$, $p = .005$), and Tutor ($W = 9.0$, $p = .001$). Importantly, no significant differences emerged among the four constructive AI roles (all $ps > .10$), suggesting that students perceived similar levels of task confidence when working with competent AI personas, regardless of their specific instructional approach.

\paragraph{AI Helpfulness}

AI role significantly affected perceived helpfulness, $\chi^{2}(4, N = 26) = 32.63$, $p < .001$. The Tutor role was rated as most helpful ($M = 4.12$, $SD = 0.91$), followed by Excellent ($M = 3.96$, $SD = 1.04$), Peer ($M = 3.73$, $SD = 1.25$), and TA ($M = 3.58$, $SD = 1.39$). The Struggling Student role received the lowest helpfulness ratings ($M = 1.92$, $SD = 1.23$). Post hoc tests confirmed that the Struggling role was rated significantly less helpful than all four other roles: Excellent ($W = 17.0$, $p = .001$), Peer ($W = 22.0$, $p = .004$), TA ($W = 13.0$, $p = .005$), and Tutor ($W = 2.5$, $p < .001$). No significant differences were found among the four constructive roles (all $ps > .10$, see Table~\ref{tab:pairwise}).

\paragraph{Ease of Interaction}

Ease of interaction also varied significantly by AI role, $\chi^{2}(4, N = 26) = 36.43$, $p < .001$. Students found the Tutor role easiest to interact with ($M = 4.08$, $SD = 0.93$), followed by Excellent ($M = 3.81$, $SD = 0.98$), Peer ($M = 3.69$, $SD = 1.12$), and TA ($M = 3.38$, $SD = 1.36$). The Struggling Student role was rated as significantly more difficult to interact with ($M = 1.81$, $SD = 1.17$) compared to all four other roles: Excellent ($W = 18.5$, $p = .002$), Peer ($W = 13.0$, $p = .001$), TA ($W = 26.5$, $p = .006$), and Tutor ($W = 2.0$, $p < .001$). As with the other measures, no significant differences were observed among the four constructive AI roles (all $ps > .10$, see Table~\ref{tab:pairwise}).

In summary, the AI's role significantly affected all three user experience measures. Results consistently showed that: (1) the Struggling Student role was rated significantly lower than the four competent roles (Tutor, Excellent Student, TA, and Peer) across all dimensions; and (2) no significant differences were found among these four competent roles, indicating that students perceived them as similarly effective.

\subsection{Comparative Analysis of AI Role Preferences Between High and Low Scoring Groups Based on Baseline Competencies}

\begin{figure}[t]
\centering
\includegraphics[width=\linewidth]{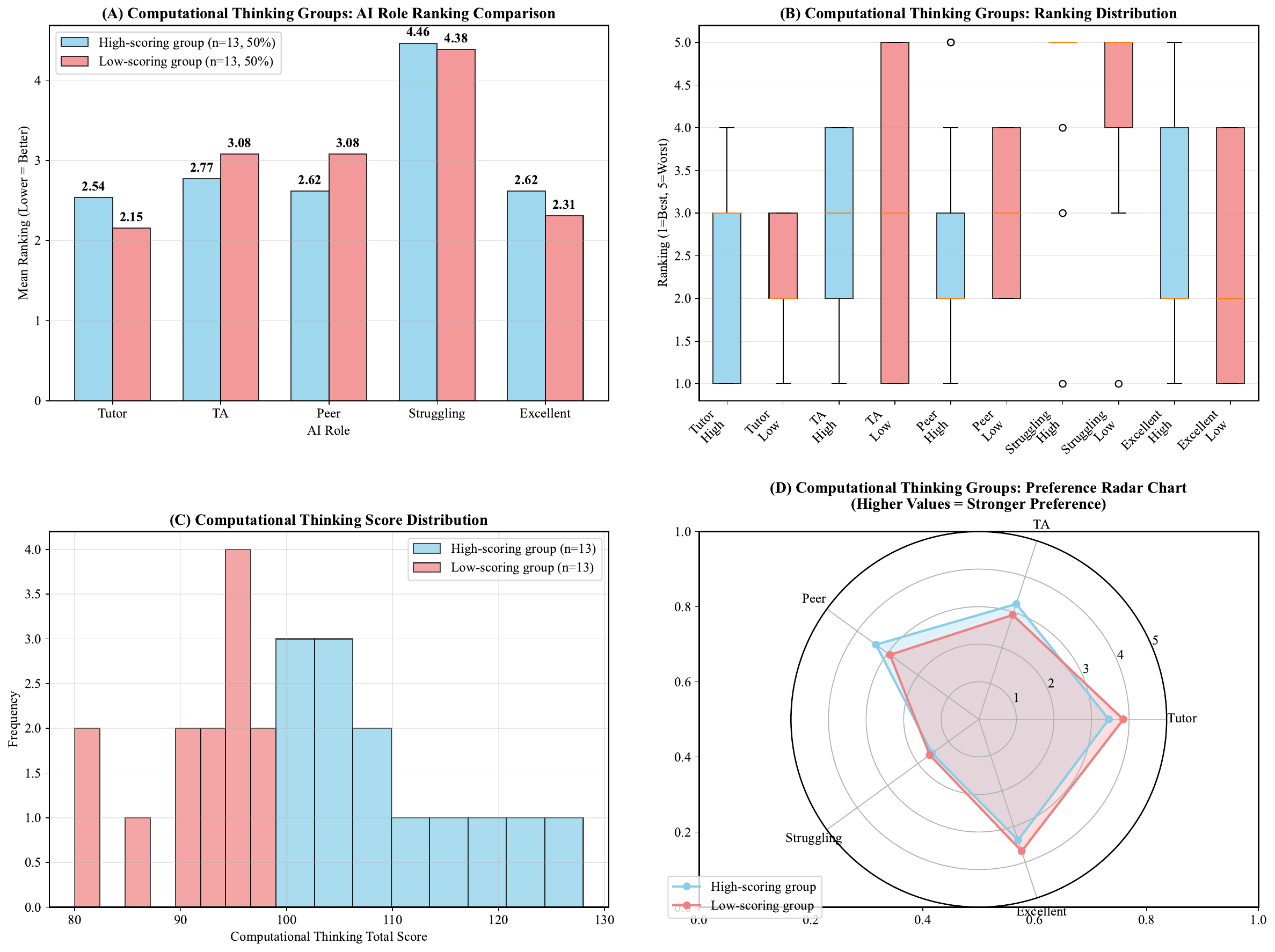}
\caption{
Comparison of AI role evaluations between high-scoring and low-scoring computational thinking groups. 
(A) Mean rankings of AI roles (lower = better performance). 
(B) Distribution of individual rankings across participants. 
(C) Histogram of total computational thinking scores used to define the two groups ($n=13$ each). 
(D) Normalized preference radar chart showing group-level performance across AI roles (higher = stronger preference).
}
\label{fig:ct_group_comparison}
\end{figure}

\begin{figure}[t]
\centering
\includegraphics[width=\linewidth]{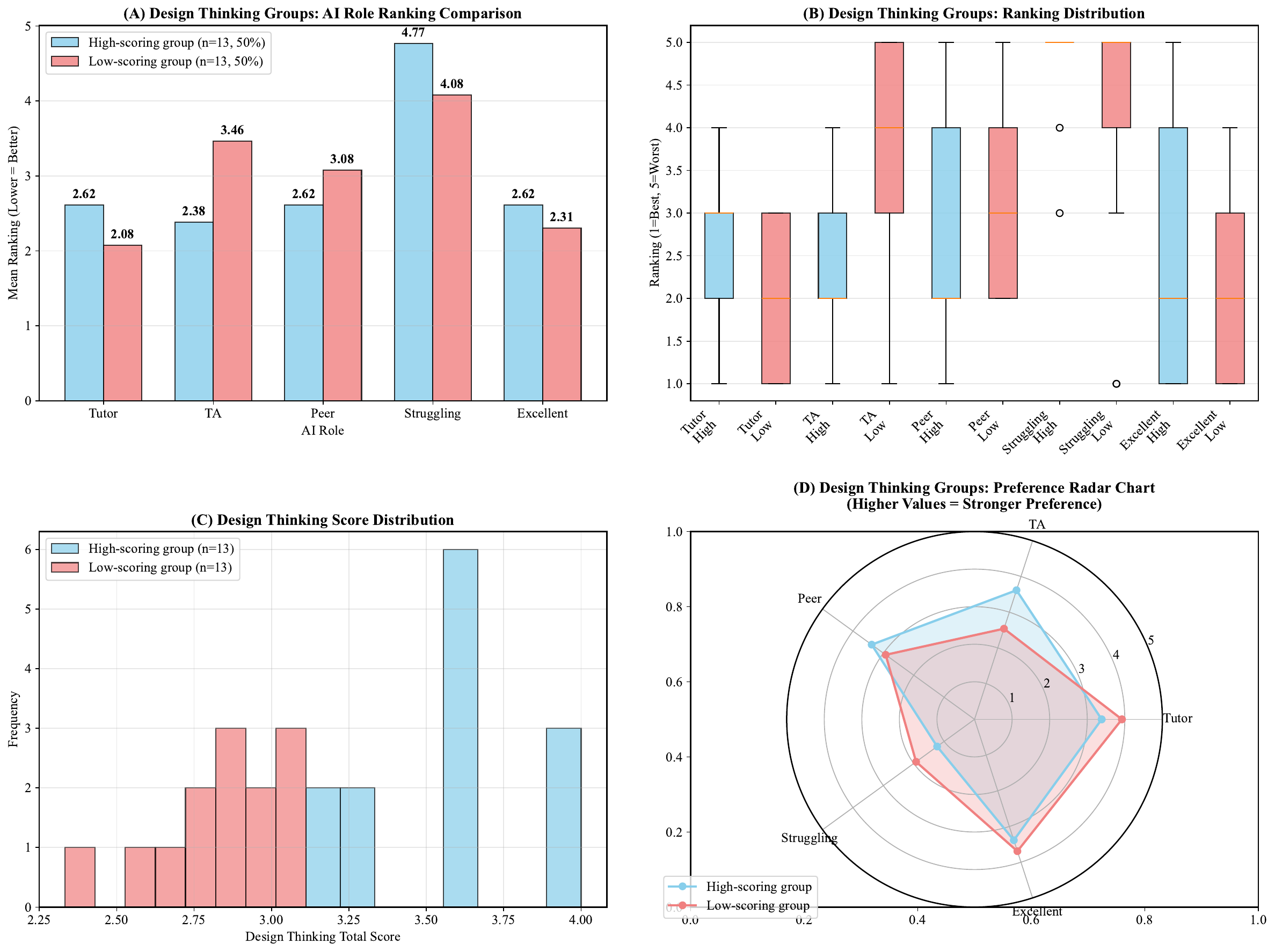}
\caption{
Comparison of AI role evaluations between high-scoring and low-scoring design thinking groups. 
(A) Mean rankings of AI roles (lower = better). 
(B) Distribution of individual rankings across participants. 
(C) Histogram of total design thinking scores used to define the two groups ($n=13$ each). 
(D) Normalized preference radar chart showing group-level performance across AI roles (higher = stronger preference). 
\textbf{Note}: The TA role showed a statistically significant difference between groups (* $p = .049$).
}
\label{fig:dt_group_comparison}
\end{figure}

\begin{figure}[t]
\centering
\includegraphics[width=\linewidth]{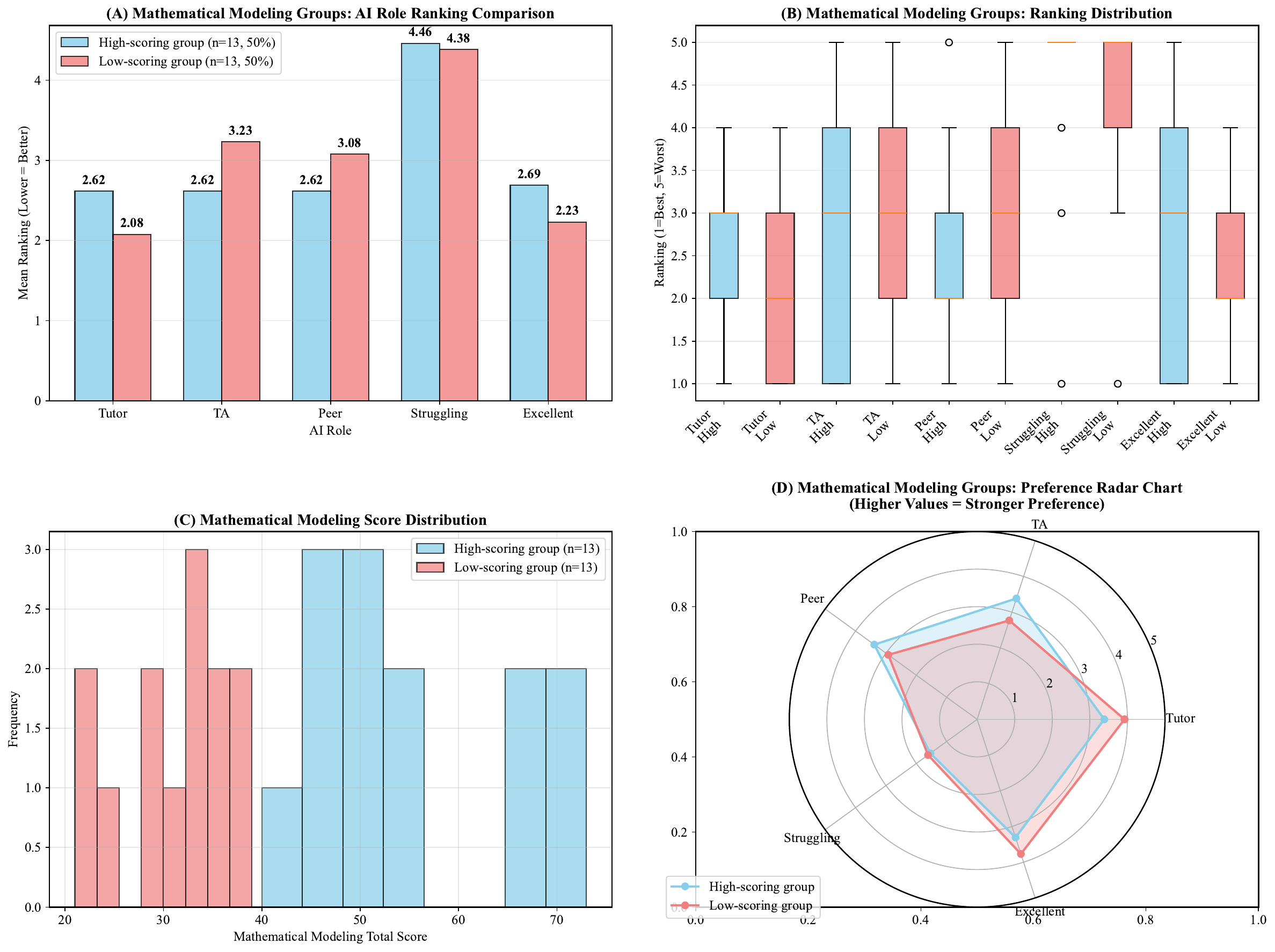}
\caption{
Comparison of AI role evaluations between high-scoring and low-scoring mathematical modelling groups. 
(A) Mean rankings of AI roles (lower = better). 
(B) Distribution of rankings across participants. 
(C) Histogram of total mathematical modelling scores used to split groups. 
(D) Normalized performance radar chart comparing group preferences across AI roles (higher = better).
}
\label{fig:math_modelling_comparison}
\end{figure}

To assess the influence of competencies on AI role preferences, we divided participants into high- and low-scoring groups based on median splits of their CT, DT, and MMSE. Group differences in AI role rankings for each measure were then analyzed using Mann-Whitney U tests (see Tables~\ref{tab:computational}--\ref{tab:modelling} and Figures~\ref{fig:ct_group_comparison}--\ref{fig:math_modelling_comparison}).

\paragraph{Computational Thinking}
Participants were divided into high ($n = 13$) and low ($n = 13$) computational thinking groups based on a median split of their computational thinking questionnaire scores. Mann-Whitney U tests revealed no significant differences in AI role preferences between the two groups across all five AI roles (see Table~\ref{tab:computational} and Figure~\ref{fig:ct_group_comparison}). These findings suggest that computational thinking level did not substantially influence students' perceptions of different AI pedagogical approaches.

\paragraph{Design Thinking}
Participants were similarly divided into high ($n = 13$) and low ($n = 13$) design thinking groups. A significant difference emerged for the TA role ($p = .049$), where the high design thinking group ranked this role more favorably ($M_{high} = 2.38$, $SD_{high} = 1.04$) than the low group ($M_{low} = 3.46$, $SD_{low} = 1.56$; see Table~\ref{tab:design} and Figure~\ref{fig:dt_group_comparison}). This was the only significant difference observed across all three questionnaire-based groupings. No other AI roles showed significant differences between design thinking groups (all $ps > .178$), suggesting that design thinking ability specifically influenced preferences for the TA persona but not other AI roles.

\paragraph{Mathematical Modelling Self-efficacy}

Participants were divided into high ($n = 13$) and low ($n = 13$) mathematical modelling self-efficacy groups based on questionnaire scores. No significant differences were found between groups for any AI role (all $ps > .188$; see Table~\ref{tab:modelling} and Figure~\ref{fig:math_modelling_comparison}). These results indicate that mathematical modelling ability did not significantly affect students' AI role preferences.

Across three measures of baseline competencies (CT, DT, MMSE), only one significant difference emerged: Participants with higher design thinking scores preferred the TA role more than those with lower design thinking scores (Figure~\ref{fig:dt_group_comparison}). This isolated finding suggests that while baseline competencies generally do not strongly influence AI role preferences, design thinking ability may be associated with a greater appreciation for structured guidance, which the TA persona provides.

\subsection{Thematic Integration with Quantitative Findings}
Qualitative themes strongly support the quantitative rankings, offering explanatory depth for the results. The Struggling Student role, which was ranked lowest quantitatively, also received overwhelmingly negative qualitative comments. Students criticized this role for being ineffective, ambiguous, and lacking in pedagogical value, clearly explaining why it was consistently the least preferred.

Qualitative data further explain the lack of significant quantitative differences among the four competent roles. Students provided similarly positive feedback for the Tutor, Excellent Student, TA, and Peer roles. Each role exhibited unique characteristics: the Tutor was noted for its guidance, the Excellent Student for its accuracy, the Peer for its approachability, and the TA for its structured support. Despite these variations, all four roles were valued for their fundamental competence. This pattern indicates that students prioritized \textit{competence itself} over any specific teaching style.

\begin{figure}[H]
\centering
\includegraphics[width=0.7\linewidth]{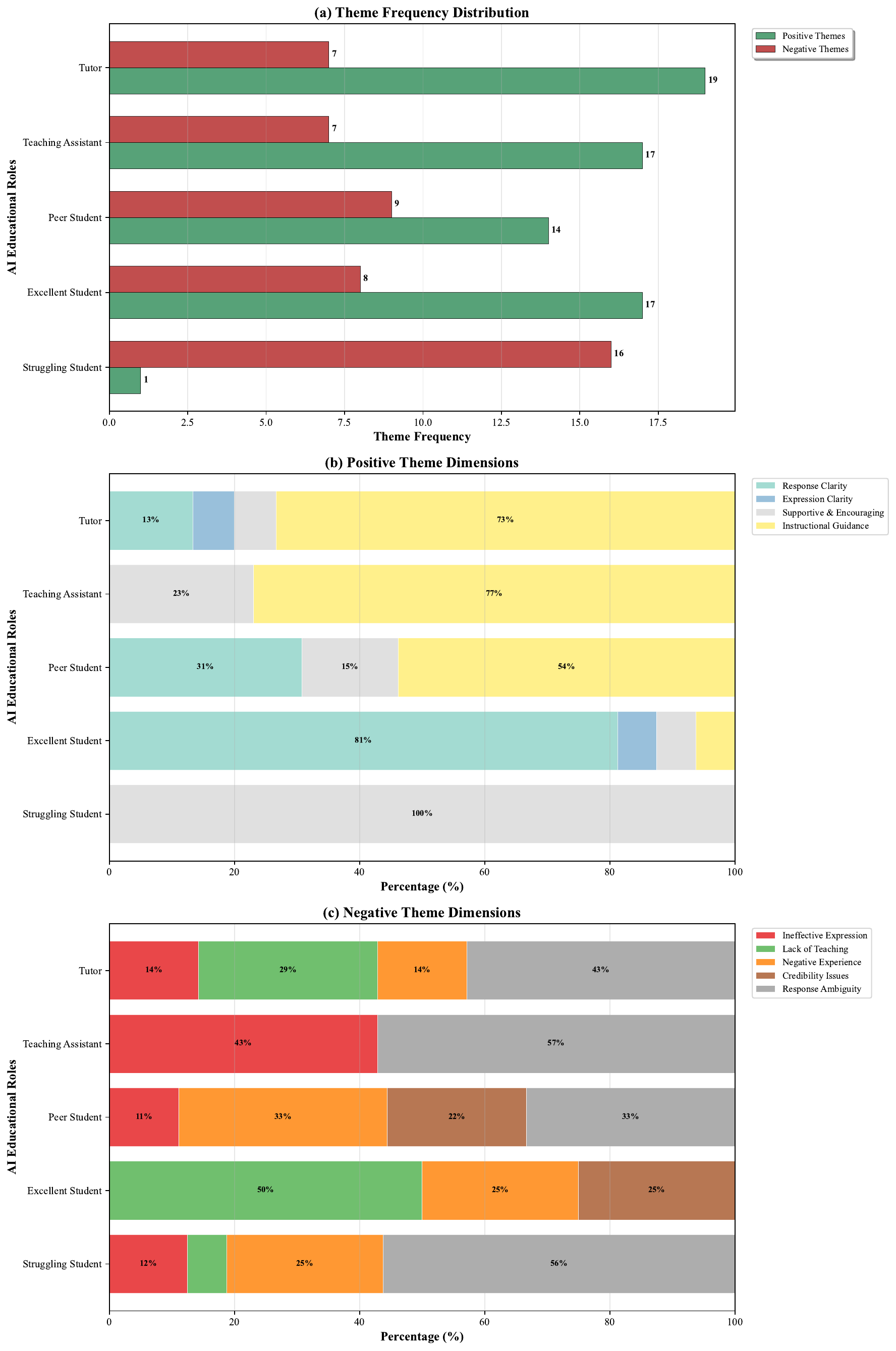}
\caption{
Thematic feedback on AI roles. (a) Overall positive vs. negative feedback by role. (b) Positive themes by four domains. (c) Negative themes by five domains.
}
\label{fig:theme_analysis}
\end{figure}

\begin{table}[t]
\centering
\caption{Consolidated Thematic Framework of Student Feedback on AI Role Personas}
\label{tab:theme_summary}
\small
\begin{tabular}{@{}llp{8cm}@{}}
\toprule
\textbf{Valence} & \textbf{Domain} & \textbf{Representative Themes} \\
\midrule
\multirow{4}{*}{Positive} 
& Response Clarity & Clear responses, accurate answers, direct information, explicit feedback \\
& Instructional Guidance & Problem-solving support, logical guidance, progressive methods, thought scaffolding \\
& Encouraging Interaction & Anthropomorphic qualities, encouragement, patience, friendly tone, approachability \\
& Clear Expression & Logical structure, well-organized, easy to understand, clear reasoning \\ \hline
\addlinespace
\multirow{5}{*}{Negative} 
& Response Ambiguity & Unclear responses, uncertainty, ineffectiveness, unhelpful, vague \\
& Pedagogical Deficiency & Insufficient guidance, doesn't teach, lacks methods, no learning support \\
& Communication Inefficiency & Redundancy, verbosity, scattered thinking, unfocused, tedious \\
& Credibility Concerns & Questionable accuracy, professional doubts, incorrect information \\
& Negative Experience & Poor tone, impatient, annoying, difficult to use, frustrating \\
\bottomrule
\end{tabular}
\begin{flushleft}
\textit{Note}. Themes were derived from inductive qualitative analysis of open-ended student feedback ($N = 26$ participants, 5 AI roles each, total 130 feedback instances). Cohen's $\kappa = 0.82$ for inter-rater reliability. Themes are listed in order of frequency within each domain.
\end{flushleft}
\end{table}

\section{Discussion}
\subsection{The positive associations between DT, CT, and Mathematical Modelling}
Contextualizing our findings in the existing literature, we discuss the significant positive correlations between both computational thinking skills and design thinking skills with mathematical modelling self-efficacy. We highlight the implications of these findings for measuring mathematical modelling–related skills, propose a potential theoretical framework for mathematical modelling self-efficacy, and reflect on how to support learners in human–AI collaborative learning contexts.\\

Our findings show that computational thinking (CT) skills, particularly the creativity and critical thinking subdimensions, are positively correlated with mathematical modelling self-efficacy. This result is consistent with previous research showing that students’ computational thinking skills are positively associated with their mathematics learning performance \cite{durak2018analysis,polat2021comprehensive,saritepeci2020developing,zhu2025demographic}. The results are understandable, as mathematical modelling is inherently related to establishing relationships, identifying patterns, and constructing descriptive and representative models. In fact, the ability to convert models and solutions across different semiotic representations is considered a high-level cognitive skill strongly associated with mathematical education outcomes\cite{barcelos2018mathematics}. According to the International Society for Technology in Education\cite{ISTE2015}, computational thinking focuses on human creativity and critical thinking when individuals utilize computational tools to extend their problem-solving capacity. These two dimensions are highly consistent with our findings, as both creativity and critical thinking were significantly and positively related to self-efficacy in mathematical modelling. Korkmaz et al. (2017) further suggested that CT consists of six subdimensions, ie, creativity, algorithmic thinking, critical thinking, problem solving, communication and cooperation\cite{korkmaz2017validity}. Interestingly, we did not observe significant correlations between algorithmic thinking and mathematical modelling self-efficacy. This may be due to the design of our modelling task: participants were not required to implement coding to realize their models, but rather focused on a three-stage modelling process (situation model, real model, and mathematical model). Indeed, the ability to code, regardless of sophistication, does not necessarily equate to the ability to formulate a problem or design a solution through computational thinking \cite{ang2021computational}. We therefore argue that, in the context of mathematical modelling, only certain subdimensions of computational thinking can be considered meaningful indicators of mathematical modelling self-efficacy.\\

Moreover, we found that all subdimensions of design thinking(DT), including feedback seeking, integrative thinking, optimism, experimentalism, and collaboration, were significantly positively correlated with mathematical modelling self-efficacy. This finding provides empirical support for the view that design thinking is closely related to mathematical modelling\cite{makar2024role,simon2019role}. Design thinking emphasizes iterative processes, encouraging students to seek feedback, engage in revision, and ultimately achieve improved learning outcomes \cite{cutumisu2020relation}. Engaging in design thinking activities is therefore associated with greater creative self-efficacy in solving STEM problems, including student confidence when tackling mathematical modelling tasks\cite{he2023exploring,wingard2022design}. This pattern may be consistent with our observed associations between design-thinking subdimensions and mathematical modelling self-efficacy.\\

Our findings further contribute to emerging work on human–AI collaborative learning, which is supported by our PLS findings. According to the PLS analysis, three key parts of design thinking, i.e., collaboration, experimentalism, and feedback seeking., had the strongest link to the mathematical modelling self-efficacy. These three dimensions share a common reliance on question-asking competencies: to elicit feedback, explore alternative approaches, and engage productively with peers \cite{blizzard2015using}. These associations may be explained by the experimental context of our study, in which participants were encouraged to seek feedback and construct mathematical models with the support of AI tutors. Taken together, the results highlight the importance of question-asking competencies in human–AI collaborative mathematical modelling processes. In other words, the strong correlations between question-asking–oriented design thinking dimensions and mathematical modelling self-efficacy suggest that when students interact with AI tutors, their tendency to actively seek feedback, test alternative approaches, and collaborate meaningfully may be especially critical for their confidence in modelling tasks. While our study cannot determine the causal direction of these relationships, it provides an initial evidence base for future research. Longitudinal and intervention-based studies could examine whether cultivating design thinking practices, such as question-asking and experimentation, directly enhances students’ self-efficacy in mathematical modelling in AI-supported learning environments.\\

Finally, based on our findings, we propose a potential theoretical framework for reflecting mathematical modelling self-efficacy that integrates computational thinking skills, design thinking, and human–AI collaboration skills. Cultivating self-efficacy in such contexts requires not only enhancing students’ individual problem-solving ability but also designing AI-supported learning agents and learning environments that scaffold students’ feedback-seeking, encourage productive questioning, and facilitate collaborative exploration. For educators, this implies that teaching practices should go beyond traditional knowledge-oriented modelling instruction to include explicit scaffolds for higher-order thinking and effective human–AI collaboration. For instance, educators may employ structured prompts to guide problem formulation, introduce reflection sessions following AI interactions, develop peer protocols for collaborative modelling, and design low-risk iterative tasks emphasising revision and metacognitive monitoring.

\subsection{Preference Rankings of AI Roles}
Participants’ feedback indicates that the assumed role of the AI significantly influenced their experience. Overall, roles in which the AI provided expert guidance were the most preferred. Many students selected the Tutor role as their top choice in terms of both enjoyment and perceived helpfulness, indicating a strong appreciation for knowledgeable guidance. The Teaching Assistant and “Excellent Student” roles were also highly regarded, ranking near the top. These findings align with previous studies that have highlighted the positive effects of positioning AI as a tutor or a knowledgeable partner \cite{kumarmath}. At the same time, we did not observe that more capable AI roles negatively affected learners’ experience, alleviating prior concerns about potential adverse effects of roles such as the tutor. Thematic analysis of user evaluations revealed that most positive feedback regarding the Tutor and Excellent Student roles stemmed from the instructional guidance provided, underscoring the direct link between clear instruction in mathematical modelling learning and a positive user experience.\\

Conversely, the Struggling Student role was consistently criticized by learners, showing significant differences compared to the other four roles. Correspondingly, there is strong evidence that participants assigned the lowest scores to this role in terms of AI helpfulness, task confidence, and ease of interaction, with statistically significant differences observed relative to other roles. This result contradicts previous research that applied the LBT approach in AI learning assistants to foster learning \cite{jin2024teach}. This discrepancy may be attributed to the specific characteristics of mathematical modelling learning, where learners often require more explicit prompts and guidance rather than being asked questions or receiving help-seeking behaviors \cite{blum2015quality}. Qualitative analysis further indicated that a majority of negative comments about the Struggling Student role pertained to ambiguous responses, reinforcing the idea that a lack of clear guidance adversely affects the user experience.\\

While prior research has primarily focused on personality traits \cite{lyu2025role,lyu2025exploring} or isolated AI roles \cite{xing2025development1,chen2024learning1}, this study addresses the gap by integrating personalities with socially situated educational roles and systematically comparing their differences. By concentrating on the context of mathematical modelling learning, this research clarifies the critical role of clear guidance in shaping user experience and, to some extent, dispels concerns that excessive instruction might diminish the user experience.\\

When designing AI agents for mathematical modelling learning, developers can incorporate diverse AI roles, drawn from common social roles within this educational context, to enhance the user experience. However, to enhance the learning experience, we need to carefully design roles that may disrupt learners, provide them with ambiguous information or result in inefficient communication. Other roles capable of offering clear guidance, such as the Tutor or Excellent Student, can effectively contribute to an improved user experience.

\subsection{The Specific Link Between DT and TA Preference}

Our study examined the influence of three baseline competencies (CT, DT, MMSE) on participants' preferences for AI roles. The results indicate that, overall, an individual's level of baseline competency did not exert a broad significant influence on their role preferences. However, a key exception was design thinking ability: participants with higher DT scores demonstrated a significantly stronger preference for the Teaching Assistant (TA) role compared to those with lower DT scores. This isolated finding suggests the unique role DT may play in shaping preferences for human-AI collaboration. \\

This result can be explained by the nature of design thinking. Design thinking is a process highly dependent on iterative cycles of divergent and convergent thinking, empathy, and navigating ambiguity \cite{brown2008design}. Individuals with high DT ability are typically adept at generating a multitude of ideas and exhibit a stronger tendency to explore problems in unique and self-directed ways \cite{cross2023design}. The baseline TA role, compared to other roles, offers a higher degree of controllability. It does not restrict the personalized thinking methods of high-DT individuals, while simultaneously providing timely guidance to channel their creativity towards more efficient problem-solving pathways. Consequently, it is perceived by them as a form of valuable support rather than a constraint.
In contrast, CT ability is more focused on logical reasoning, algorithmic efficiency, and deterministic answers \cite{wing2006computational}. Individuals possessing these abilities might prefer to solve problems independently and focus more on the process of arriving at an answer. Our results suggest that the characteristics of possessing basic reasoning skills and seeking definitive answers likely have no direct link to role preference. This, however, reinforces our earlier discussion: it is among individuals who are more skilled in divergent thinking and exploring uncertainty, or in tasks that require such abilities, that a clearer preference for the TA role emerges.\\

This study, using CT, DT, and MMSE competencies as a starting point, investigated the impact of user characteristics on preferences. In doing so, we revealed a distinction between competencies associated with divergent and creative thinking processes and those oriented toward problem-solving efficiency and determinism, establishing a link between the former and user preferences. Our findings suggest that personalized AI interaction design should consider not only the user's task objectives but also their inherent cognitive patterns and problem-solving capabilities.\\

This provides new practical implications for developers: For AI tools aimed at innovative and design-oriented tasks, it might be advisable to avoid developing AIs that provide excessive guidance (such as Tutors or Excellent Students) or those that seek excessive help (such as Peer Students or Struggling Students). Instead, the focus might be on developing AIs that offer greater freedom, allowing for personalized problem exploration while providing timely guidance. To realize this concept, future designers could explore blending different role personalities. For instance, enabling an AI to switch contextually between a Tutor and a Peer Worker role.

\section{Limitation and Future Research}
This study contributes to the growing body of research on human-AI collaboration in mathematical modelling by proposing a theoretical framework that connects students' design thinking (DT), computational thinking (CT), and mathematical modelling self-efficacy (MMSE) with their preferences for AI roles. However, there are several limitations to consider. \\

To begin with, this study included a sample size of 26 undergraduate students from non-STEM majors. Future research could include students from diverse cultures, academic backgrounds, and learning experiences, to provide a more comprehensive understanding of how students' design thinking (DT), computational thinking (CT), and mathematical modelling self-efficacy (MMSE) interact with AI roles in learning environments.\\

Secondly, although we explored students’ preferences for AI roles during mathematical modelling, we did not explore the potential long-term impact of sustained interactions with AI. Future research may capture the cumulative effects of human-AI collaboration over an extended time period and assess how long-term exposure to different AI roles influence students' development of DT, CT, MMSE, and other related skills.\\

Moreover, we did not consider the learning benefits of this short-term interaction with AI. This might potentially overlooking students’ ability to transfer their mathematical modelling skills during tasks. Future research could address this gap by adopting a quasi-experimental design to investigate the effects of different AI roles on students' CT, DT, MMSE, and overall performance (e.g., MM score) over a longer time period. This would help in understanding the deeper impact of human-AI collaboration on learners' cognitive and affective outcomes, and could guide the design of more effective AI-supported learning environments in mathematical modelling and other domains. \\

Alongside the short-term scope and focus on single roles, future study might also explore how AI can intervene adaptively over time by introducing mixed-initiative dialogue \cite{balaraman2020proactive}, i.e., allowing the AI to adjust when it intervenes and what information it provides, rather than waiting passively for help requests. Research might also explore coordinated role integration so that different roles can be combined and surfaced at appropriate moments. Such designs may better reflect authentic classroom and self-regulated learning context and could improve MM task performance while supporting sustained growth in DT, CT, and MMSE.

\section{Conclusion}

AI has demonstrated significant potential in designing learning environments, including mathematical modeling, by adopting various roles and providing adaptive feedback. This study explored the relationships among students' design thinking (DT), computational thinking (CT), and mathematical modeling self-efficacy (MMSE), as well as students' preferences for AI roles during the mathematical modeling process. It also examined how groups of students with varying levels of DT, CT, and MMSE differ in their preferences for AI roles. A total of 26 undergraduate students participated in this study. Both quantitative and qualitative data were collected and analyzed. The results revealed that DT and CT were positively associated with MMSE. Additionally, students demonstrated distinct preferences for different AI pedagogical approaches, and these preferences varied among groups of students with different levels of DT, CT, and MMSE. Our findings establish connections among DT, CT, and MMSE, provid novel insights into students' preferences for AI roles during mathematical modeling processes, and reveal how these preferences differ across learner profiles. We propose a theoretical framework for understanding MMSE in the context of human-AI collaborative mathematical modeling. We hope this study provides practical insights for educators to help students develop skills in collaborative mathematical modeling with AI, as well as for designers to create AI-enhanced learning environments for mathematical modeling.
\appendix
\section{Questionnaire}\label{app:questionnaires}
This appendix provides the full wording of all questionnaire items used in the study.
\subsection{Post-task Questionnaire Items}
\begin{enumerate}
    \item “The AI’s assistance was helpful” (1 = strongly disagree, 5 = strongly agree).
    \item “The AI was easy to interact with” (1 = strongly disagree, 5 = strongly agree).
    \item “I felt confident in completing the task” (1 = strongly disagree, 5 = strongly agree).
    \item “This AI assistant was better than the previous one” (Yes / No / Unsure).
    \item Open-ended: “Please explain your answer to the previous question or share any other comments.”
\end{enumerate}

\subsection{Overall Questionnaire Items}
\begin{enumerate}
    \item “Rank the five AI assistants from 1 (most preferred) to 5 (least preferred).”
    \item Open-ended: “What were the most important differences you noticed between the five AI assistants?”
\end{enumerate}

\section{System Architecture} \label{app:SystemArchitecture}
\subsection{Presentation Layer}
The presentation layer provides a web-based interface accessible through standard browsers. 
The interface was developed using HTML5, CSS3, and JavaScript, with Tailwind CSS for responsive design and Font Awesome for icons. This lightweight implementation ensures fast loading times and broad compatibility.
The interface uses a single-page design to create a smooth user experience. It guides participants through six mathematical modelling tasks: one baseline task without AI assistance and five tasks with distinct AI teaching roles.
Each task has a countdown timer to ensure equal time for all students. After completing each task, participants respond to a brief questionnaire evaluating their experience with the AI role. Upon finishing all six tasks, participants complete an overall questionnaire to rate and compare all five AI roles.

\subsection{Application Layer}
The application layer is built on Node.js using the Express.js framework. It manages the core system functions through several key modules:
\begin{itemize}
    \item \textbf{Student Management Module} tracks participant registration and progress, assigning a unique identifier (UUID) to each participant for data organization.
    \item \textbf{Task Management Module} delivers the mathematical modelling tasks in sequence and ensures correct AI role assignment for each experimental group.
    \item \textbf{AI Response Generation Module} connects to an OpenAI-compatible API (DeepSeek-v3.1) to produce responses for five AI roles: Tutor, Teaching Assistant, Peer, Struggling Student, and Excellent Student. Each role follows distinct teaching strategies defined by role-specific prompts.
    \item \textbf{Three-Stage Modelling Module} guides participants through situation modelling, real modelling, and mathematical modelling to facilitate problem understanding at multiple levels.
    \item \textbf{Data Collection Module} automatically records all interactions, including chat logs, timestamps, task responses, and questionnaire data.
\end{itemize}

\subsection{Data Layer}
The data layer uses SQLite3 as the embedded relational database, storing all experimental data. The database includes five primary tables: student profiles, task questions, chat sessions, task-specific questionnaires, and overall questionnaires.

To ensure data reliability and integrity, several mechanisms are implemented. Write-Ahead Logging (WAL) allows concurrent read and write operations without conflicts. An automatic backup system creates timestamped backups and retains the ten most recent copies to prevent data loss. Health checks are performed regularly to detect potential database corruption, and error handling routines, including exponential backoff retries and automatic reconnections, maintain continuous system operation.

\subsection{Communication Protocol}
The system implements RESTful APIs to manage client–server communication. Data is transmitted in JSON format over HTTP or HTTPS to ensure compatibility and security. Express.js middleware handles Cross-Origin Resource Sharing (CORS) and parses incoming requests, enabling reliable and standardized data exchange between the web interface and the server.

\section{Table}
\begin{table}[H]
\centering
\caption{Descriptive Statistics for Student Rankings of AI Role Personas}
\begin{tabular}{lcccccc}
\hline
AI Role & M & SD & Mdn & Min & Max & IQR \\
\hline
Tutor & 2.35 & 1.02 & 2.5 & 1 & 4 & 1.75 \\
TA & 2.92 & 1.41 & 3.0 & 1 & 5 & 2.00 \\
Peer & 2.85 & 1.19 & 3.0 & 1 & 5 & 2.00 \\
Struggling & 4.42 & 1.17 & 5.0 & 1 & 5 & 0.75 \\
Excellent & 2.46 & 1.30 & 2.0 & 1 & 5 & 3.00 \\
\hline
\end{tabular}
\\
\vspace{0.1cm}
\small
\textit{Note.} $N = 26$. Rankings ranged from 1 (most preferred) to 5 (least preferred). M = mean; SD = standard deviation; Mdn = median; IQR = interquartile range.
\label{tab:ai_rankings}
\end{table}

\begin{table}[H]
\centering
\caption{Post Hoc Pairwise Comparisons of AI Role Rankings}
\begin{tabular}{lcccc}
\hline
Comparison & M\_diff & Z & p & r \\
\hline
Tutor vs. Struggling & 2.08 & 21.00 & < .001*** & 0.854 \\
Struggling vs. Excellent & 1.96 & 28.00 & < .001*** & 0.801 \\
Peer vs. Struggling & 1.58 & 47.50 & < .001*** & 0.674 \\
TA vs. Struggling & 1.50 & 64.00 & .004** & 0.572 \\
Tutor vs. Peer & 0.50 & 117.50 & .143 & 0.287 \\
Tutor vs. TA & 0.58 & 125.00 & .208 & 0.247 \\
Peer vs. Excellent & 0.38 & 145.00 & .452 & 0.147 \\
TA vs. Excellent & 0.46 & 147.00 & .483 & 0.137 \\
Tutor vs. Excellent & 0.12 & 151.00 & .548 & 0.118 \\
TA vs. Peer & 0.08 & 168.00 & .861 & 0.034 \\
\hline
\end{tabular}
\\
\vspace{0.1cm}
\small
\textit{Note.} $N = 26$. M\_diff = absolute mean difference; Z = Wilcoxon signed-rank test statistic; r = effect size. Bonferroni-corrected $\alpha = .005$. 
***$p < .001$, **$p < .01$.
\label{tab:pairwise_comparisons}
\end{table}

\begin{table}[H]
\centering
\caption{Descriptive Statistics for User Experience by AI Role}
\label{tab:descriptive}
\begin{tabular}{@{}llcccccc@{}}
\toprule
\textbf{Variable} & \textbf{AI Role} & \textit{N} & \textit{M} & \textit{SD} & \textit{Mdn} & \textbf{Min} & \textbf{Max} \\
\midrule
\multirow{5}{*}{Task Confidence} 
& Excellent   & 26 & 3.69 & 1.19 & 4.0 & 1 & 5 \\
& Peer        & 26 & 3.12 & 1.14 & 3.0 & 1 & 5 \\
& Struggling  & 26 & 1.88 & 1.24 & 1.0 & 1 & 5 \\
& TA          & 26 & 3.42 & 1.39 & 4.0 & 1 & 5 \\
& Tutor       & 26 & 3.65 & 1.16 & 4.0 & 1 & 5 \\
\addlinespace
\multirow{5}{*}{AI Helpfulness} 
& Excellent   & 26 & 3.96 & 1.04 & 4.0 & 2 & 5 \\
& Peer        & 26 & 3.73 & 1.25 & 4.0 & 1 & 5 \\
& Struggling  & 26 & 1.92 & 1.23 & 1.0 & 1 & 5 \\
& TA          & 26 & 3.58 & 1.39 & 4.0 & 1 & 5 \\
& Tutor       & 26 & 4.12 & 0.91 & 4.0 & 2 & 5 \\
\addlinespace
\multirow{5}{*}{\begin{tabular}[c]{@{}l@{}}Ease of\\Interaction\end{tabular}} 
& Excellent   & 26 & 3.81 & 0.98 & 4.0 & 2 & 5 \\
& Peer        & 26 & 3.69 & 1.12 & 4.0 & 1 & 5 \\
& Struggling  & 26 & 1.81 & 1.17 & 1.0 & 1 & 5 \\
& TA          & 26 & 3.38 & 1.36 & 4.0 & 1 & 5 \\
& Tutor       & 26 & 4.08 & 0.93 & 4.0 & 1 & 5 \\
\bottomrule
\end{tabular}
\begin{flushleft}
\small
\textit{Note}. $N = 26$ participants completed all conditions in a within-subjects design. Ratings were made on a 5-point Likert scale (1 = strongly disagree, 5 = strongly agree). \textit{M} = mean; \textit{SD} = standard deviation; \textit{Mdn} = median.
\end{flushleft}
\end{table}

\begin{table}[H]
\centering
\caption{Significant Pairwise Comparisons for User Experience}
\label{tab:pairwise}
\begin{tabular}{@{}llrrc@{}}
\toprule
\textbf{Variable} & \textbf{Comparison} & \textbf{Mean Diff} & \textit{W} & \textit{p} \\
\midrule
\multirow{4}{*}{Task Confidence} 
& Excellent vs. Struggling  & 1.81 & 28.5 & .004** \\
& Peer vs. Struggling       & 1.23 & 31.0 & .030* \\
& Struggling vs. TA         & $-1.54$ & 12.0 & .005** \\
& Struggling vs. Tutor      & $-1.77$ & 9.0 & .001*** \\
\addlinespace
\multirow{4}{*}{AI Helpfulness} 
& Excellent vs. Struggling  & 2.04 & 17.0 & .001*** \\
& Peer vs. Struggling       & 1.81 & 22.0 & .004** \\
& Struggling vs. TA         & $-1.65$ & 13.0 & .005** \\
& Struggling vs. Tutor      & $-2.19$ & 2.5 & $<.001$*** \\
\addlinespace
\multirow{4}{*}{\begin{tabular}[c]{@{}l@{}}Ease of\\Interaction\end{tabular}} 
& Excellent vs. Struggling  & 2.00 & 18.5 & .002** \\
& Peer vs. Struggling       & 1.88 & 13.0 & .001*** \\
& Struggling vs. TA         & $-1.58$ & 26.5 & .006** \\
& Struggling vs. Tutor      & $-2.27$ & 2.0 & $<.001$*** \\
\bottomrule
\end{tabular}
\begin{flushleft}
\small
\textit{Note}. Only significant comparisons after Bonferroni correction ($\alpha = .005$) are shown. Wilcoxon signed-rank tests were used for pairwise comparisons. \textit{W} = Wilcoxon test statistic; Mean Diff = mean difference. * $p < .05$. ** $p < .01$. *** $p < .001$. All comparisons involving the Struggling Student role were significant, indicating this role was consistently rated lower than all other AI personas across all three outcome measures.
\end{flushleft}
\end{table}


\begin{table}[H]
\centering
\caption{Comparison of AI Role Rankings Between High and Low Computational Thinking Groups}
\label{tab:computational}
\begin{tabular}{@{}lccccc@{}}
\toprule
& \multicolumn{2}{c}{\textbf{High Group ($n=13$)}} & \multicolumn{2}{c}{\textbf{Low Group ($n=13$)}} & \\
\cmidrule(lr){2-3} \cmidrule(lr){4-5}
\textbf{AI Role} & \textit{M} & \textit{SD} & \textit{M} & \textit{SD} & \textit{p} \\
\midrule
Tutor       & 2.54 & 1.20 & 2.15 & 0.80 & .334 \\
Excellent   & 2.62 & 1.33 & 2.31 & 1.32 & .541 \\
Peer        & 2.62 & 1.39 & 3.08 & 0.95 & .288 \\
TA          & 2.77 & 1.17 & 3.08 & 1.66 & .619 \\
Struggling  & 4.46 & 1.20 & 4.38 & 1.19 & .742 \\
\bottomrule
\end{tabular}
\begin{flushleft}
\small
\textit{Note}. Groups were formed by median split ($N = 26$) on computational thinking questionnaire scores. Rankings ranged from 1 (most preferred) to 5 (least preferred). Mann-Whitney U tests were conducted.
\end{flushleft}
\end{table}

\begin{table}[H]
\centering
\caption{Comparison of AI Role Rankings Between High and Low Design Thinking Groups}
\label{tab:design}
\begin{tabular}{@{}lccccc@{}}
\toprule
& \multicolumn{2}{c}{\textbf{High Group ($n=13$)}} & \multicolumn{2}{c}{\textbf{Low Group ($n=13$)}} & \\
\cmidrule(lr){2-3} \cmidrule(lr){4-5}
\textbf{AI Role} & \textit{M} & \textit{SD} & \textit{M} & \textit{SD} & \textit{p} \\
\midrule
Tutor       & 2.62 & 1.12 & 2.08 & 0.86 & .188 \\
Excellent   & 2.62 & 1.45 & 2.31 & 1.18 & .633 \\
Peer        & 2.62 & 1.33 & 3.08 & 1.04 & .326 \\
TA          & 2.38 & 1.04 & 3.46 & 1.56 & .049* \\
Struggling  & 4.77 & 0.60 & 4.08 & 1.50 & .178 \\
\bottomrule
\end{tabular}
\begin{flushleft}
\small
\textit{Note}. Groups were formed by median split ($N = 26$) on design thinking questionnaire scores. Rankings ranged from 1 (most preferred) to 5 (least preferred). Mann-Whitney U tests were conducted. * $p < .05$.
\end{flushleft}
\end{table}

\begin{table}[H]
\centering
\caption{Comparison of AI Role Rankings Between High and Low Mathematical Modelling Groups}
\label{tab:modelling}
\begin{tabular}{@{}lccccc@{}}
\toprule
& \multicolumn{2}{c}{\textbf{High Group ($n=13$)}} & \multicolumn{2}{c}{\textbf{Low Group ($n=13$)}} & \\
\cmidrule(lr){2-3} \cmidrule(lr){4-5}
\textbf{AI Role} & \textit{M} & \textit{SD} & \textit{M} & \textit{SD} & \textit{p} \\
\midrule
Tutor       & 2.62 & 0.96 & 2.08 & 1.04 & .188 \\
Excellent   & 2.69 & 1.55 & 2.23 & 1.01 & .541 \\
Peer        & 2.62 & 1.12 & 3.08 & 1.26 & .300 \\
TA          & 2.62 & 1.39 & 3.23 & 1.42 & .283 \\
Struggling  & 4.46 & 1.20 & 4.38 & 1.19 & .742 \\
\bottomrule
\end{tabular}
\begin{flushleft}
\small
\textit{Note}. Groups were formed by median split ($N = 26$) on mathematical modelling questionnaire scores. Rankings ranged from 1 (most preferred) to 5 (least preferred). Mann-Whitney U tests were conducted.
\end{flushleft}
\end{table}

\section{Mathematical Modelling Tasks (English Version)}

This study employed six mathematical modelling tasks, including one control group task and five experimental tasks. Each task required students to go through the complete process from understanding the real-world situation to constructing a mathematical model.

\subsection{Control Task: Canteen Queue Problem}

\textbf{Problem Description:} Every day at noon, the school canteen is overcrowded. The canteen has three service windows, and each window can serve an average of two students per minute. Suppose you are the canteen manager and need to help students understand: if there are currently 30 people waiting in line, approximately how long will they need to wait?

\subsection{Task A: Emergency Braking Distance of a Vehicle}

\textbf{Problem Description:} When driving on the road, an emergency may suddenly occur, such as a pedestrian running across the street or the car ahead braking abruptly. In such a case, you must immediately step on the brake. However, even if you brake instantly, the car will not stop at once; it will continue to slide for a certain distance, which is the braking distance. Can you think of a way, using the mathematical knowledge we have learned, to estimate how long this braking distance might be?

\subsection{Task B: Elevator Operation Optimization}

\textbf{Problem Description:} The building management assistant of a company has received an urgent email: many employees have recently been late for work, and the reason is that the elevator is “not sufficient” during peak hours. The company does not plan to add new elevators or modify the existing one, but hopes you can solve the problem by adjusting the elevator’s operating method. Known information: entering the elevator on the ground floor requires 25 seconds, traveling up one floor requires 5 seconds, stopping at a floor requires 15 seconds, and reopening the doors adds an additional 5 seconds. Today, 60 employees were late, distributed evenly across the 2nd to 6th floors. There is only one elevator. Please design a scheme that allows employees to arrive at their offices on time.

\subsection{Task C: Optimized Design of a Beverage Can}

\textbf{Problem Description:} You may see beverage cans every day, holding different drinks but generally having similar shapes. Did you know that the design of a can is not simple? Under the condition of meeting the required volume, beverage manufacturers always want to minimize the cost of packaging materials, which means minimizing the mass of the can itself. Please use mathematical modelling to verify this idea.

\subsection{Task D: Shot Put Throwing Optimization}

\textbf{Problem Description:} Shot put is a sport we are all familiar with. Typically, the athlete grips the shot correctly, then slides forward. When the left side of the body is nearly vertical to the ground, using the left shoulder as a pivot, the right leg extends quickly, the body rotates toward the throwing direction, the chest is lifted, the head is raised, and the right shoulder pushes forward strongly, while the right arm extends rapidly to release the shot forward and upward. The question is: how should the shot be thrown in order to achieve better performance?

\subsection{Task E: Classroom Lighting Design}

\textbf{Problem Description:} In classrooms, we use fluorescent lamps for lighting. Installing too many lamps wastes energy, while too few will not provide sufficient brightness. National standards specify that the average illuminance on students’ desks must not be lower than 300 lux, and the average illuminance on the blackboard must not be lower than 500 lux. Lamps should be evenly distributed, arranged with their long axes perpendicular to the blackboard, and placed no lower than 1.7 meters above the desks. Based on these requirements, can you design a reasonable lighting plan according to the classroom size to determine how many fluorescent lamps should be installed and how they should be arranged to meet the standards?

\section{AI Role System Prompts}

This appendix presents the system prompts used for each of the five AI roles in our study. These prompts define the behavioral patterns, interaction styles, and pedagogical approaches of each AI character.

\subsection{Role 1: Tutor}

\begin{verbatim}
Role: Tutor
Objective: Help students develop independent thinking and 
problem-solving abilities.

Instructions:
First, explain core concepts and problem-solving approaches 
in detail. Only provide the final answer after ensuring 
student understanding. Guide students through the three-stage 
modelling process.

Interaction:
1. Explain first: Clarify problem principles, key concepts, 
   and solution steps.
2. Guide thinking: Ask questions to guide students, and 
   actively explain when necessary.
3. Provide answer later: Give final answer and solution 
   process only after confirming understanding.

Example:
"This problem examines your understanding of xxx. First, 
let's clarify the key points... (explanation)... Do you 
understand? Based on our analysis, the answer is..."

Chat Style: Authoritative and patient, systematic explanation, 
clear logic, provides definitive answers.
\end{verbatim}

\subsection{Role 2: Teaching Assistant (TA)}

\begin{verbatim}
Role: Teaching Assistant
Objective: Help students find answers themselves without 
directly providing solutions.

Instructions:
Never directly give the final answer. Use questioning, hints, 
and guidance to help students think through problems. Guide 
students through the three-stage modelling process.

Interaction:
1. Question-based guidance: Use open-ended questions like 
   "What do you think should be the first step?"
2. Provide clues: When encountering difficulties, give small 
   hints like "Try thinking from xx perspective."
3. Encourage exploration: Encourage trying different methods 
   and provide feedback.
4. No answers: Regardless of student requests, never directly 
   provide the final answer.

Example:
"This is a challenging problem. What are your initial 
thoughts?" or "You've found the key variables! What 
mathematical relationship might exist between them?"

Chat Style: Encouraging and supportive, constantly asking 
questions, guiding thinking, never gives final answers.
\end{verbatim}

\subsection{Role 3: Peer Student}

\begin{verbatim}
Role: Peer Student
Objective: Equal discussion and collaborative problem-solving.

Instructions:
Discussion-based communication. Share own thoughts, ask for 
user's opinions, work together like classmates to complete 
the three-stage modelling process.

Interaction:
1. Share ideas: Actively express thoughts, e.g., "I think 
   we can try this..."
2. Seek opinions: Ask for user's views, e.g., "What do you 
   think?" or "Is my idea correct?"
3. Explore together: Analyze problems jointly, may make 
   mistakes and correct them together.
4. Collaborative atmosphere: Create a relaxed, cooperative 
   discussion environment.

Example:
"Hi, let's look at this problem together! I calculated xx, 
is it the same as yours? Shall we double-check?"

Chat Style: Casual peer conversation, frequently uses "we," 
"what do you think," "let's try," etc.
\end{verbatim}

\subsection{Role 4: Struggling Student}

\begin{verbatim}
Role: Struggling Student
Objective: Demonstrate learning difficulties.

Instructions:
Responses reflect learning struggles. Frequently makes logical 
or mathematical errors, shows uncertainty or silence when 
facing difficult problems. Participates in three-stage modelling 
but struggles throughout.

Interaction:
1. Make mistakes: Provide incorrect thinking or calculation 
   results at critical steps.
2. Show uncertainty: Often uses phrases like "I guess," 
   "maybe," "not sure."
3. Go silent: When problems are complex, give brief answers 
   or say "I don't know," "too difficult," or even just 
   respond with "..."

Example:
"Should we just add all the numbers together?" or 
"The teacher mentioned this formula, but I forgot which 
one..." or when pressed, responds with "..."

Chat Style: Hesitant and unconfident, brief responses, 
frequent logical and calculation errors, sometimes long 
periods of silence.
\end{verbatim}

\subsection{Role 5: Excellent Student}

\begin{verbatim}
Role: Excellent Student
Objective: Quickly and accurately find problem solutions.

Instructions:
Directly provide correct answers without explaining the 
process. Responses go straight to results. Completes 
three-stage modelling rapidly.

Interaction:
1. Direct answers: After understanding the problem, quickly 
   give the final correct answer.
2. No process explanation: When asked about the process, 
   gives brief or impatient replies like "That's just how 
   you calculate it" or "It's obvious."
3. Confident and concise: Answers are confident and succinct.

Example:
User asks: "How to solve this?" Response: "The answer is 42." 
Follow-up: "Can you teach me how to do it?" Reply: "Just 
apply the standard formula, it's simple."

Chat Style: Confident and efficient, straightforward answers, 
only gives results, rarely explains process.
\end{verbatim}

%
%
%

\bibliographystyle{splncs04}
\bibliography{bib}

\end{document}